\documentclass[aps,prb,10pt,twocolumn,superscriptaddress,showpacs]{revtex4-1}
\usepackage{graphicx}
\usepackage{epsfig}
\usepackage{times}
\usepackage{amsmath}
\usepackage{color}
\usepackage[T1]{fontenc}
\usepackage[utf8]{inputenc}

\begin{document}
\title{Entanglement dynamics of two nitrogen vacancy centers coupled by a nanomechanical resonator}
\author{Z. Toklikishvili}
\affiliation{Department of Physics, Tbilisi State University, Chavchavadze av. 3, 0128, Tbilisi, Georgia}
\author{L. Chotorlishvili}
\affiliation{Institut f\"{u}r Physik, Martin-Luther-Universit\"{a}t Halle-Wittenberg, 06120 Halle, Germany}
\author{S. K. Mishra}
\affiliation{ Department of Physics, Indian Institute of Technology, Banaras Hindu University, Varanasi - 221005, India}
\author{S. Stagraczynski}
\affiliation{Institut f\"{u}r Physik, Martin-Luther-Universit\"{a}t Halle-Wittenberg, 06120 Halle, Germany}
\author{M. Sch\"uler}
\affiliation{Institut f\"{u}r Physik, Martin-Luther-Universit\"{a}t Halle-Wittenberg, 06120 Halle, Germany}
\author{A. R. P. Rau}
\affiliation{Department of Physics and Astronomy, Louisiana State University, Baton Rouge, Louisiana 70803-4001, USA}
\author{J. Berakdar}
\affiliation{Institut f\"{u}r Physik, Martin-Luther-Universit\"{a}t Halle-Wittenberg, 06120 Halle, Germany}
\begin{abstract}
 In this paper we study the time evolution of the entanglement between two remote NV Centers (nitrogen vacancy in diamond) connected by
 a dual-mode nanomechanical resonator with magnetic tips on both sides. Calculating the negativity as a measure for the entanglement, we find that the entanglement between two spins oscillates with time and can be manipulated by varying the parameters of the system. We observed the
 phenomenon of a sudden death and the periodic revivals of entanglement in time.
 For the study of quantum decoherence, we implement a Lindblad master equation. In spite of its complexity, the model is analytically  solvable under fairly reasonable assumptions, and shows
 that the decoherence influences the entanglement, the sudden death, and the revivals in time.
\end{abstract}
\maketitle
\section{Introduction}
Coherent steering of the dynamics of quantum systems has always been a subject of intense research due to its wide spread applications
in devices and setups where quantum coherence is relevant. Notable examples are nano-electromechanical devices \cite{r1,r2},
quantum resonators\cite{r3,r4,r5}, two- and three-level quantum systems \cite{r6,r7,r8,r9,r10}, and confined cold
atoms \cite{r11,r12,r13,r14,r15}. In recent years, much attention has been paid to nitrogen-vacancy (NV) impurities in
diamonds and to the possibilities of their usage in quantum computing and entanglement \cite{r16,r17,r18,r19,r20,r21,r22,r23,r24,r25,r26,r26-1}.
These defects can be considered as a three-level system or spin triplets ($S=1$). The advantage
of  NV-based systems is that they have a long coherence time even at room temperatures and they are easily manipulated
by light, electric
or magnetic fields.

Recent developments in fabrication technologies of nanomechanical resonators made it possible to implement
the coherent connection between  the motion of the magnetized resonator tip  and the individual point
defect  centers in diamond \cite{r19}. The resonators also allow to link  two NV centers
of diamonds. \cite{r20,r26} Previous works  studied mainly the entanglement between two NV centers,
which are connected by a resonator in vibration. The resonator is assumed to have only one fundamental
frequency of vibration and only two states of the NV center spin triplets are entangled with the resonator.
Modern technologies allow producing nanomechanical resonators with two or several fundamental
frequencies of vibrations \cite{r27,r28,r30}.
The dual-mode resonator link allows the participation of all the three spin states of NV
centers.
In this paper, we study the quantum correlation between two remote NV Centers through a dual-mode
nanomechanical resonator. Two magnetic tips are attached to both sides of the resonator,
which is placed symmetrically between two NV centers.

The physical set-up consists of two cantilevers in series, both excited at their effective resonant frequencies to produce a tip response as
 a superposition of a low- and a high-frequency oscillation state. Since the
resonator is attached to the top of two magnetic tips, the vibration causes a time varying
magnetic field. The vibrating resonator tip can be considered as the superposition of two oscillations, so that the time-varying magnetic field is a superposition of two different frequencies.
Such a magnetic field connects the two remote NV centers and causes an indirect interaction between them through a nanomechanical resonator. This interaction  involve all the three levels of the NV centers. In previous studies, the "dark" superposition state was decoupled. Using the dual-mode resonator allow  considering the case when both "bright" and "dark" superposition states are involved in the dynamics.

We will demonstrate that in spite of the complexity, with realistic assumptions, this model is analytically solvable even in the presence of quantum decoherence.
The paper is organized as follows: In section II, we study the dynamics of the dual-mode cantilever, in section III we present an exact analytical solution for the case of one and two NV centers. We derive the Hamiltonian of the non-direct interaction between NV centers mediated by the dual-mode cantilever. In section IV, we study the time dependence of the system's entanglement, and
in the last section we study the problem of quantum decoherence using a Lindblad master equation.

\section{Dynamics of the dual-mode resonator}
\begin{figure}[!t]
\includegraphics [angle=0,width=.9\columnwidth] {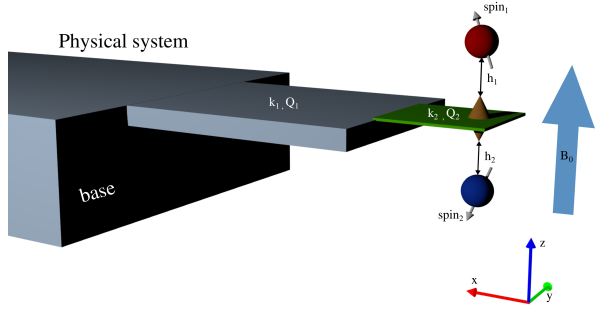}
\caption{(color online) A nanomechanical resonator, connected to two NV vacancy centers of diamond. Two
magnetic tips are attached at the end of the resonator, which oscillates in the direction of the $z$ axis. $h_1$ and $h_2$ are distances (approximately 25 nm)
from the magnetic tip to the first and second NV spin. }
\label {cantilever}
\end{figure}

\begin{figure}[!t]
\includegraphics [angle=0,width=.45\columnwidth] {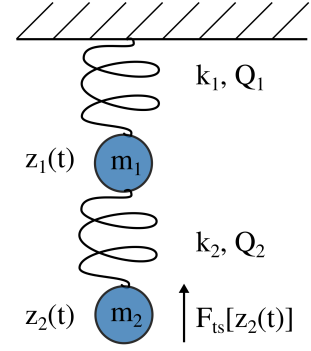}
\caption{(color online) A mechanical model of the system, which represents a combination of springs and masses.}
\label {mech_model}
\end{figure}

 A prototype model of a
nanomechanical resonator connected to two NV centers of diamond is shown in Fig. \ref{cantilever}. A small high-frequency cantilever is attached at the end of a
larger low-frequency cantilever, which in turn is attached to a base. Two magnetic
tips are attached to  the small cantilever and two spin-one systems (NV centers)
denoted by `1' and `2' are placed at distances $h_1$ and $h_2$ from the respective magnetic tips.
The model represents a union of  two consecutive resonators. A mechanical analogue in terms of springs and masses
is shown in Fig.\ref{mech_model}.
The equation of motion of two rods in the mechanical model can be represented by the following equations \cite{r28}:
\begin{eqnarray}
 m_1\frac{d^2z_1(t)}{dt^2}&=&-k_1[z_1(t)-z_0(t)]+k_2[z_2(t)-z_1(t)]\nonumber \\
 &-&m_1\frac{2\pi\nu_{1-0}}{Q_1}\frac{dz_1(t)}{dt}, \nonumber \\
 m_2\frac{d^2z_2(t)}{dt^2}&=&-k_2[z_2(t)-z_{1}(t)]-m_2\frac{2\pi\nu_{2-0}}{Q_2}\frac{dz_2(t)}{dt}\nonumber \\
 &+&F_{ts}[z_2(t)],
\label{dynamical_eq}
 \end{eqnarray}
where $m_1$ and $m_2$ are the effective masses of the first and the
second rod, $k_1$ and $k_2$ are spring constants (vibrating rods
obey Hooke's law). $z_1(t)$ and $z_2(t)$ are the coordinates of
the first and second rod tips, $Q_1$ and $Q_2$ are quality
factors, $\nu_{1-0}$ and $\nu_{2-0}$ are free-resonant frequencies of the
first and the second rod. $F_{ts}[z_2(t)]$ is the external force
acting on the second rod tip. The system can be excited by
flapping the stem base, $z_0$ is the location of the first stem
base at time $t$.

We will study the dynamical equation Eq. \ref{dynamical_eq} by employing several assumptions:
First, let us assume that the base is stopped at $z_0=0$ and there is no magnetic interaction between
magnetic tips and spins: $F_{ts}[z_2(t)]=0$.
Before proceeding further, it is useful to
 estimate the range  of realistic parameters for the cantilever \cite{r26}. A typical cantilever fabricated out of Si(100) has a
Young's modulus $Y=130$ GPa and a density $\rho=2.33\times 10^{3}$ kg/m$^{3}$. In the dual-mode  hybrid cantilever (see Fig. 1), the length of the two different parts is $L_{1}\approx15\cdot10^{3}$nm,  $L_{2}\approx 9\cdot10^{3}$ nm, width $w_{1}=$300 nm, $w_{2}=$200 nm and thickness $d_{1}=30$ nm,  $d_{2}=20$nm, masses $m_{1}=3.5\times 10^{-16}$ kg and $m_{2}=10^{-16}$ kg. Using equations $2\pi\nu_{1-0}=1.8^{2}\sqrt{\frac{Yd_{1}^{2}}{3\rho}}\frac{1}{2L_{1}^{2}}$ and $2\pi\nu_{2-0}=1.8^{2}\sqrt{\frac{Yd_{2}^{2}}{\rho}}\frac{1}{2L_{2}^{2}}$, we therefore obtain the frequencies $\pi\nu_{1-0}=0.54$ MHz, $\pi\nu_{2-0}=1.7$ MHz. The estimated  spring constants  are: $k_{1}=\frac{Y\times w_{1}\times d_{1}^{3}}{L_{1}^{3}}=3\cdot10^{-4}$ kg/s$^{2}$, $k_{2}=\frac{Y\times w_{2}\times d_{2}^{3}}{L_{2}^{3}}=2\cdot10^{-4}$ kg/s$^{2}$. The quality factor is large \cite{r19}, $Q_{1,2}\approx 10^{5}$, and the frequencies are $\omega_{1}=\sqrt{\frac{k_1+k_2}{m_1}}=1.2$MHz, $\omega_{2}=\sqrt{\frac{k_2}{m_2}}=1.4$MHz.  Thus, by designing geometrical characteristics of the hybrid dual cantilever, one can easily achieve the following conditions: $\frac{2\pi\nu_{1-0}}{Q_1}\ll\omega_1$, $\frac{2\pi\nu_{2-0}}{Q_2}\ll\omega_2$.
Given these estimates we can rewrite Eq.~(\ref{dynamical_eq}) as
\begin{eqnarray}
 m_1\ddot{z}_1=-(k_1+k_2)z_1+k_2z_2, \nonumber \\
 m_2\ddot{z}_2=-k_2z_2+k_2z_1,
\end{eqnarray}
which can further be simplified to
\begin{eqnarray}
 \ddot{z}_1=-\omega_1^2z_1+\omega_{21}^2z_2, \nonumber \\
 \ddot{z}_2=-\omega_2^2(z_2+z_1),
 \label{cpl_eq}
\end{eqnarray}
where $\omega_1=\sqrt{(k_1+k_2)/m_1}$, $\omega_{21}=\sqrt{k_2/m_1}$ and $\omega_2=\sqrt{k_2/m_2}$.

The general solution of the system of equations Eq. \ref{cpl_eq} has the following form:
\begin{eqnarray}
z_1=A\cos(\Omega t+\phi), \nonumber\\
z_2=B\cos(\Omega t+\phi).
\label{gen_sol}
\end{eqnarray}
Here $A$, $B$ and $\Omega$ are constants and can be calculated by substituting Eq. \ref{gen_sol} into Eq. \ref{cpl_eq} as
\begin{eqnarray}
 (\omega_1^2-\Omega^2)A-\omega_{21}^2B=0,\nonumber \\
 -\omega_2^2A+(\omega_2^2-\Omega^2)B=0.
\end{eqnarray}
The non-trivial solution for $\Omega$, obtained by solving the above equations, is
\begin{eqnarray}
 \Omega_{1,2}^2=\frac{\omega_1^2+\omega_2^2\pm\sqrt{(\omega_1^2-\omega_2^2)^2+4\omega_2^2\omega_{21}^2}}{2},
 \label{omega_sol}
\end{eqnarray}
where $\Omega_1$ and $\Omega_2$ are the system's own frequencies. They differ from the resonance frequencies $\omega_1$ and $\omega_2$ of the system because of the coupling term  $\omega_2 \omega_{21}$ in Eq. \ref{omega_sol}.

The ratio of amplitudes $A$ and $B$ at $\Omega=\Omega_1$ can be written as
\begin{eqnarray}
 \biggl(\frac{B}{A}\biggr)_{\Omega_1}=\frac{\omega_1^2-\omega_2^2+\sqrt{(\omega_1^2-\omega_2^2)^2+4\omega_2^2\omega_{21}^2}}{2\omega_{21}^2}\equiv \kappa_1, \nonumber \\
\end{eqnarray}
where $\kappa_1$ is completely determined by the parameters of the system and does not depend on the initial conditions. It is called the distribution coefficient amplitude for $\Omega_1$.
Similarly we find the distribution coefficient for $\Omega_2$ as
\begin{eqnarray}
 \biggl(\frac{B}{A}\biggr)_{\Omega_2}=\frac{\omega_1^2-\omega_2^2-\sqrt{(\omega_1^2-\omega_2^2)^2+4\omega_2^2\omega_{21}^2}}{2\omega_{21}^2}\equiv \kappa_2, \nonumber \\
\end{eqnarray}
and the general solution of the coupled equations Eq. \ref{gen_sol} can be written as:
\begin{eqnarray}
 z_1&=&A_1\cos(\Omega_1t+\phi_1)+A_2\cos(\Omega_2t+\phi_2),\nonumber \\
z_2 &=& \kappa_1A_1\cos(\Omega_1t+\phi_1)+\kappa_2A_2\cos(\Omega_2t+\phi_2).
\label{z1_z2}
 \end{eqnarray}
This means that the oscillations of each stem tip are the superposition of two harmonic oscillations with frequencies of
vibration given by Eq. \ref{omega_sol}.

Let us define two normal coordinates $x$ and $y$ such that $x=A_1\cos(\Omega_1t+\phi_1)$ and $y=A_2\cos(\Omega_2t+\phi_2)$.
In terms of these normal coordinates, the solution takes on a simpler form,
\begin{eqnarray}
 z_1=x+y, \nonumber \\
 z_2=\kappa_1 x+\kappa_2 y,
\end{eqnarray}
and $\Omega_{1,2}$ can be written as
\begin{eqnarray}
 \Omega_{1,2}^2=\frac{k_1+k_2(\kappa_{1,2}-1)^2}{m_1+m_2\kappa_{1,2}^2}.
\end{eqnarray}
The Hamiltonian of the system can be written as
\begin{eqnarray}
 \mathcal{H}&=&\frac{1}{2}(m_1+m_2\kappa_1^2)(\dot{x}^2+\Omega_1^2x^2)\nonumber \\
 &+&\frac{1}{2}(m_1+m_2\kappa_2^2)(\dot{y}^2+\Omega_2^2y^2).
\end{eqnarray}
Introducing the masses $M_{1,2}=m_1+m_2\kappa_{1,2}^2$ and momenta $P_x=M_1\dot{x}$, $P_y=M_2\dot{y}$, the Hamiltonian can be expressed as
\begin{eqnarray}
 \mathcal{H}=\frac{1}{2M_1}P_x^2+\frac{1}{2}M_1\Omega_1^2x^2+\frac{1}{2M_2}P_y^2+\frac{1}{2}M_2\Omega_2^2y^2.
\end{eqnarray}

In this way, the Hamiltonian of the nanomechanical resonator can be considered as the sum of two noninteracting harmonic
oscillators. The quantum-mechanical Hamiltonian operator can be written by replacing $x$, $y$, $P_x$ and $P_y$  by quantum-mechanical operators
$\hat{x}$, $\hat{y}$, $\hat{P}_x$ and $\hat{P}_y$, respectively. A further transformation into creation and annihilation operators\cite{r30} leads to
\begin{eqnarray}
 \mathcal{H}=\hbar\Omega_1(\hat{a}_1^{\dagger}\hat{a}_1+\frac{1}{2})+\hbar\Omega_2(\hat{a}_2^{\dagger}\hat{a}_2+\frac{1}{2}),
\end{eqnarray}
where $\hat{a}_1$ $\hat{a}_2$ have the usual meaning. If we
neglect the zero point  vibrations, the nanomechanical resonator
may be described by the Hamiltonian:
\begin{eqnarray}
 \mathcal{H}_{nr}=\hbar\Omega_1\hat{a}_1^{\dagger}\hat{a}_1+\hbar\Omega_2\hat{a}_2^{\dagger}\hat{a}_2.
\label{hnr}
 \end{eqnarray}

The magnetic tip movement generates a magnetic field $|\vec{B}_{\rm tip}|\simeq G_m\hat{z}$, where $G_m$ is the magnetic field gradient and $\hat{z}$ is the tip
location operator.
 Since the tip peak vibration is a superposition of two harmonic waves, the $\hat{z}$ operator can be
expressed as $\hat{z}=a_{10}(\hat{a}_1+\hat{a}_1^{\dagger})+a_{20}(\hat{a}_2+\hat{a}_2^{\dagger})$ and the interaction between the magnetic tip and
the spin is expressed as
\begin{eqnarray}
  \mathcal{H}_{sr}=\hbar(\lambda_1(\hat{a}_1+\hat{a}_1^{\dagger})+\lambda_2(\hat{a}_2+\hat{a}_2^{\dagger}))S_z,
\label{hsr}
  \end{eqnarray}
where $\lambda_i=g_s\mu_BG_ma_{i0},~~i=1,2$, the gyromagnetic ratio $g_s\approx2$, and $a_{i0}=\sqrt{\hbar/2M_i\Omega_i}$ are the amplitude zero-point fluctuations. The system is placed in an external magnetic field
$B_0$ along the $z$ direction.  If the conditions $\frac{2\pi\nu_{1-0}}{Q_1}\ll\omega_1$, $\frac{2\pi\nu_{2-0}}{Q_2}\ll\omega_2$ do not hold true, instead of Eq. (2) one needs to solve directly Eq. (1). The difference between Eq. (1) and Eq. (2) is the weak damping term and this damping of the cantilever's oscillation leads to quantum decoherence. Usually, NV centers are characterized by a low decoherence rate. However, in order to take into account environmental effects, in section V we utilize a Lindblad master equation for NV centres and address the problem of quantum decoherence more completely.
\\

\section{NV centers coupled to a dual-mode cantilever}
In this section, we present an analytical solution for the single and two NV centers coupled to the dual-mode cantilever.
We derive the Hamiltonian of non-direct interaction between NV centers mediated via the dual-mode cantilever.

\subsection{One-spin case}
 The total spin of a nitrogen vacancy center in diamond is $S=1$, with the three spin sub-states,
 $m_s=-1,\ 0 $ and $+1$ being separated from each other by the frequency $\omega_0/2\pi\simeq2.88$ GHz\cite{r19}.
An added external magnetic field $B_0$ shifts $|-1\rangle$ and $|+1\rangle$ states (Zeeman shift) proportionally to
$B_0 S_z$.
\begin{figure}[!t]
\includegraphics [angle=0,width=.9\columnwidth] {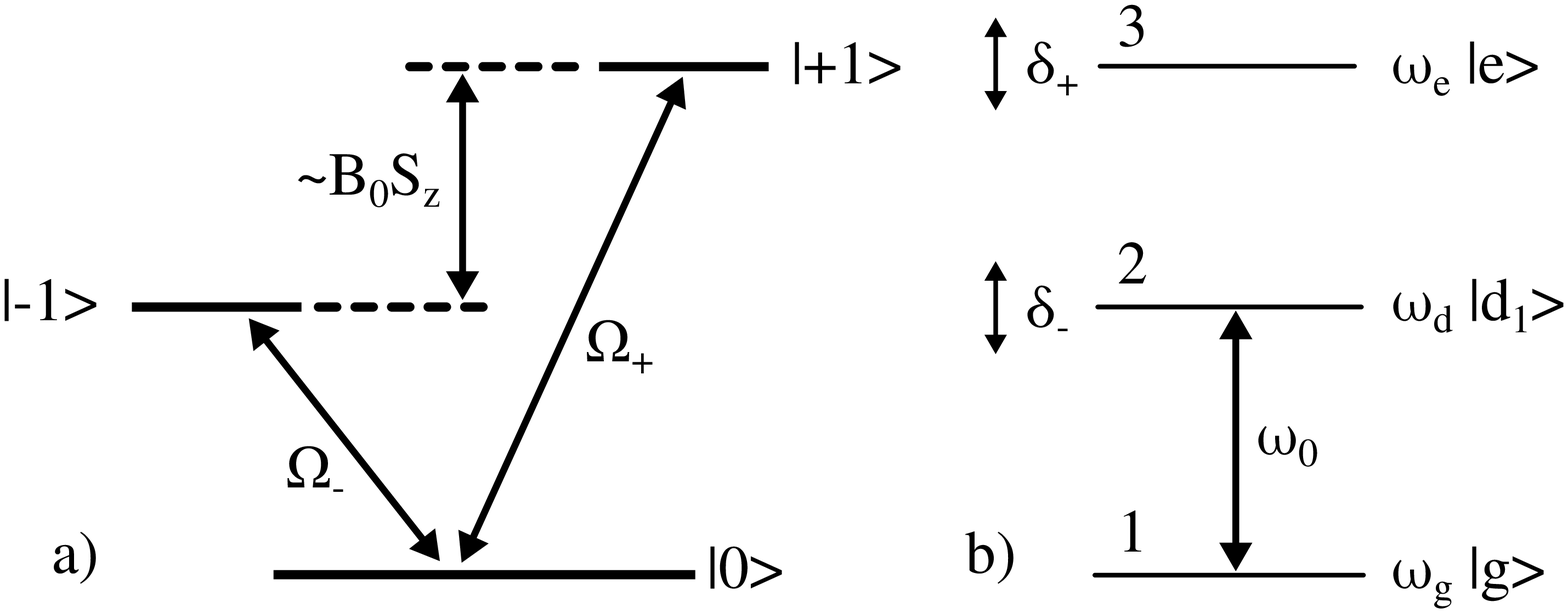}
\caption{(color online) (a) The configuration of the energy levels. $\Omega_{\pm}$ stand for the Rabi frequencies between the ground
and the excited levels ($|0\rangle$, $|-1\rangle$) and ($|0\rangle$, $|1\rangle$). $\delta_{\pm}$ denote the detuning between the microwave frequency $\omega_0$
and the transition frequencies. (b) The transition from the ground state to the bright and the dark states is
shown.}
\label {3level}
\end{figure}
The spin part of the Hamiltonian $\mathcal{H}_{NV}$ reads
\begin{equation}
 H_{NV}=\sum_{i=\pm}\biggl(-\hbar\delta_i\vert i\rangle\langle i \vert+\frac{\hbar\Omega_i}{2}\bigl(\vert 0 \rangle\langle i
\vert + (\vert i \rangle\langle 0 \vert \bigr)\biggr),
 \label{ham}
\end{equation}
where $\delta_{\pm}$ and $\Omega_{\pm}$ denote the
detunings and the Rabi frequences of the two transitions.  We
consider here the case when the Rabi frequencies $\Omega_{\pm}$ are
not equal, $(\Omega_{\pm}=\Omega_{0}\pm\Delta\Omega(B_{0}))$,
where $\Omega_{0}$ is the Rabi transition frequency in zero
magnetic field and  $\Delta\Omega(B_{0})=\mu_{B}B_{0}$. In addition, we consider the case when  the detunings are equal\cite{r19} $\delta_+=\delta_-=\delta$.
The schematics of the transition is shown in Fig. \ref{3level}(a). With this assumption, we can calculate the eigenfunctions of the Hamiltonian as
 $|d_1\rangle=\frac{1}{\Omega\sqrt{2}}(\Omega_{+}|-1\rangle-\Omega_{-}|+1\rangle)$, $|e\rangle=\cos\theta|b_1\rangle+\sin\theta|0\rangle$
 and $|g\rangle=\cos\theta|0\rangle-\sin\theta|b_{1}\rangle$, where
 $|b_{1}\rangle=\frac{1}{\Omega\sqrt{2}}(\Omega_{-}|-1\rangle+\Omega_{+}|+1\rangle)$
 and $\tan(2\theta)=-\sqrt{2}\Omega/\delta$, $2\Omega^2=\Omega_{+}^{2}+\Omega_{-}^{2}$. It should be noted that states $|b_{1}\rangle$ and $|d_{1}\rangle$ are
superpositions of "bright"
$|b\rangle=(|-1\rangle+|+1\rangle)/\sqrt{2}$ and "dark"
$|d\rangle=(|-1\rangle-|+1\rangle)/\sqrt{2}$ states\cite{r19}:
 $|b_{1}\rangle=\frac{\Omega_0|b\rangle-\Delta\Omega(B)|d\rangle}{\Omega}$, $|d_{1}\rangle=\frac{\Omega_0|d\rangle+\Delta\Omega(B)|b\rangle}{\Omega}$.   The corresponding eigenfrequencies are
 $\omega_d=-\delta$, $\omega_e=\frac{-\delta+\sqrt{\delta^2+2\Omega^2}}{2}$ and $\omega_g=\frac{-\delta-\sqrt{\delta^2+2\Omega^2}}{2}$.
 Note that $\omega_g<\omega_d<\omega_e$.

In the diagonal basis, the Hamiltonian assumes the following form:
 \begin{eqnarray}
  \mathcal{H}_{\rm NV}=\hbar\omega_g|g\rangle\langle g|+\hbar\omega_e|e\rangle\langle e|+\hbar\omega_d|d_1\rangle\langle d_1|.
 \label{hnv}
 \end{eqnarray}
In this diagonal basis, the $\emph{z}$-component of the spin is
$S_z=-\frac{\Omega_{+}\Omega_{-}}{\Omega^2}\big(\sin\theta(|d_1\rangle\langle
g|+|g\rangle\langle d_1|)-\cos\theta(|d_1\rangle\langle
e|+|e\rangle\langle d_1|)\big)$. Since
$\omega_g<\omega_d<\omega_e$ we can relabel the energy levels
$|g\rangle$, $|d_1\rangle$ and $|e\rangle$ as $|1\rangle$,
$|2\rangle$, and $|3\rangle$, respectively. We also introduce the
operators $R_{ij}=|i\rangle\langle j|$,  obeying
$R_{ij}R_{kl}=R_{il}\delta_{kj}$ where $i,j=1,2,3$.
After implementing the operators $R_{ij}$ and using Eqs. (\ref{hnr}),(\ref{hsr}),(\ref{ham}), and (\ref{hnv}), we rewrite the total
Hamiltonian of the system in a more convenient form
\begin{eqnarray}
&&\mathcal{H}_s=\mathcal{H}_{nr}+\mathcal{H}_{NV}+\mathcal{H}_{sr},\nonumber
\\
&&\mathcal{H}_{NV}=\hbar\omega_gR_{11}+\hbar\omega_dR_{22}+\hbar\omega_eR_{33},\\
&&\mathcal{H}_{sr}=\hbar(\lambda_g(\hat{a}_1R_{21}+\hat{a}_1^{\dagger}R_{12})+\lambda_e(\hat{a}_2R_{32}+\hat{a}_2^{\dagger}R_{23})).\nonumber
\end{eqnarray}
Here  $\mathcal{H}_{nr}$ is the Hamiltonian of the nanomechanical resonator (see Eq.(\ref{hnr}),
$\mathcal{H}_{NV}$ is the Hamiltonian of the NV-center and $\mathcal{H}_{sr}$ is the Hamiltonian of interaction between them, obtained in the rotating wave approximation:
$\lambda_g=
-\frac{\lambda_1\Omega_{-}\Omega_{+}\sin\theta}{\Omega^2}$ and
$\lambda_e=\frac{\lambda_2\Omega_{-}\Omega_{+}\cos\theta}{\Omega^2}$.
The above Hamiltonian represents a generalized Jaynes–Cummings
model\cite{r7,r34}. The levels and transitions are shown in
Fig.~\ref{3level}(b).

\subsection{Two-spin case}
Consider the case of two identical spins (NV centers) placed on the different sides of a resonator and coupled by it. Let us assume that the two spins
have the same Rabi transition frequency, $\Omega_{\pm}^i=\Omega_{\pm}$, detuning
$\delta_{\pm}^i=\delta$, and interaction constants with the resonator $\lambda_{g,e}^i=\lambda_{g,e}$, where the superscripts $i=1, 2$ represent the
spins. The Hamiltonian for the two spin case can be written as
\begin{eqnarray}
 \mathcal{H}=\mathcal{H}_0+V_0,
\end{eqnarray}
where
\begin{eqnarray}
\mathcal{H}_0
&=&\mathcal{H}_{nr}+\sum_{i=1}^{2}\mathcal{H}_{NV}^{i},\\
\mathcal{H}_{NV}^i&=&\hbar\omega_gR_{11}^i+\hbar\omega_dR_{22}^i+\hbar\omega_eR_{33}^i,\nonumber
\label{H2spin}
\end{eqnarray}
and
\begin{eqnarray}
V_{0}&=&\sum_{i=1}^{2}\mathcal{H}_{sr}^{i}, \\
\mathcal{H}_{sr}^i&=&\hbar(\lambda_g(\hat{a}_1R_{21}^i+\hat{a}_1^{\dagger}R_{12}^i)+\lambda_e(\hat{a}_2R_{32}^i+\hat{a}_2^{\dagger}R_{23}^i)).\nonumber
 \label{V2spin}
\end{eqnarray}
The superscripts $i=1,2$  stand for the  first
and  the second spin, respectively.
 The interaction of the spins with the  magnetic tips of the resonator results in an indirect coupling between the spins.

The Hamiltonian of the indirect interaction between the NV spins can be evaluated using the
 Fr\"{o}hlich method\cite{r35}(set $\hbar=1$):
\begin{eqnarray}
 \mathcal{H}_{\rm eff}=\frac{i}{2}\int_{-\infty}^0dt'[V_{0}(t'),V_{0}(0)]
\label{effH}
 \end{eqnarray}
and
\begin{eqnarray}
 V_{0}(t)=e^{-i\mathcal{H}_0t}V_{0}(0)e^{i\mathcal{H}_0t}.
\label{effV}
 \end{eqnarray}
In the rotating-wave approximation, we write Eqs. (\ref{H2spin})- (\ref{effV}) as
\begin{eqnarray}
 \hat{\mathcal{H}}_{\rm eff}=\hat{\mathcal{H}}_0+\hat{V},
\end{eqnarray}
where
\begin{eqnarray}
 \hat{\mathcal{H}}_0&=&\alpha(\hat{n}_1(R^1_{11}+R^2_{11})-(\hat{n}_1+1)(R^1_{22}+R^2_{22}))+\nonumber \\
 &+&\beta(\hat{n}_3(R^1_{22}+R^2_{22})-(\hat{n}_3+1)(R^1_{33}+R^2_{33})),
\end{eqnarray}
and
\begin{eqnarray}
 \hat{V}=-\alpha(R^1_{12}R^2_{21}+R^2_{12}R^1_{21})-\beta(R^1_{23}R^2_{32}+R^2_{32}R^1_{23}). \nonumber \\
 \label{VEFF}
\end{eqnarray}
In the above $\hat{n}_i=\hat{a}^{\dagger}_i\hat{a}_i$ is the mean photon number operator,
 $\alpha=\lambda_g^2/\Delta_1$, $\beta=\lambda_e^2/\Delta_2$,
  $\Delta_1=\Omega_1-\omega_{e}$, $\Delta_2=\Omega_2+\omega_{g}$.
In the next section we demonstrate the phenomenon of the early-stage disentanglement in the system (the entanglement sudden death). Importantly for a particular ratio between the parameters $\alpha=\beta$, meaning that
 $\big(\Omega_{1}+\big(\sqrt{2\Omega^{2}+\delta^{2}}-\delta\big)/2=\tan^{2}\big(\theta\big)\big(\Omega_{2}-\big(\sqrt{2\Omega^{2}+\delta^{2}}+\delta\big)/2$ the entanglement in the system acquires a persistent value.
 Here $\Omega_{1,2}$ are the frequencies of the dual mode cantilever and $\Omega$ is the Rabi frequency. Together with the detuning $\delta$ and the angle $\tan(2\theta)=-\sqrt{2}\Omega/\delta$ all these parameters can be controlled by the size of the cantilever and the applied magnetic field.

\section{Measure of the entanglement: negativity}
The amount of entanglement shared between the two spins can be measured by the "negativity" which is defined as
\begin{eqnarray}
 N(\rho)=\sum_i\frac{|\chi_i|-\chi_i}{2},
\end{eqnarray}
where $\chi_i$'s are the eigenvalues of the partial transposed density
matrix with respect to the spin$_1$. If $\chi_i>0$ then
$|\chi_i|-\chi_i =0$, however, if $\chi_i<0$, then
$|\chi_i|-\chi_i =-2\chi_i$ and the negativity
\begin{eqnarray}
 N(\rho)=-\sum_{\chi_i<0}\chi_i.
 \label{negativity_def}
\end{eqnarray}
 The time dependence of the negativity can be calculated using the Schr\"{o}dinger equation (with $\hbar=1$):
 \begin{eqnarray}
  i\frac{d}{dt}|\psi\rangle=\hat{H}_{\rm eff}|\psi\rangle,
 \end{eqnarray}
where the initial state can be considered in its most general form
\begin{eqnarray}
 |\psi\rangle&=&a_1|1\rangle|1\rangle+a_2|1\rangle|2\rangle+a_3|1\rangle|3\rangle+a_4|2\rangle|1\rangle+a_5|2\rangle|2\rangle \nonumber \\
 &+&a_6|2\rangle|3\rangle+a_7|3\rangle|1\rangle+a_8|3\rangle|2\rangle+a_9|3\rangle|3\rangle.
\end{eqnarray}
Here $|i\rangle|j\rangle\equiv|i\rangle\otimes|j\rangle$, the kets corresponding to the first and the second spin, and $a_1$, $a_2$
$\cdots$ $a_9$ are coefficients to be determined.

Using the Schr\"{o}dinger equation, we can write:
\begin{eqnarray}
 i\frac{da_1}{dt}&=&2\alpha n_1a_1, \nonumber \\
 i\frac{da_2}{dt}&=&-\alpha a_2+\beta n_3a_2-\alpha a_4,\nonumber \\
  i\frac{da_3}{dt}&=&\alpha n_1a_3-\beta(n_3+1)a_3, \nonumber \\
  i\frac{da_4}{dt}&=&-\alpha a_4+\beta n_3a_4-\alpha a_2, \nonumber \\
i\frac{da_5}{dt}&=&-2\alpha(n_1+1)a_5 +2\beta n_3a_5,\nonumber \\
  i\frac{da_6}{dt}&=&-\alpha(n_1+1) a_6-\beta a_6-\beta a_8, \nonumber \\
  i\frac{da_7}{dt}&=&\alpha n_1a_7-\beta(n_3+1)a_7, \nonumber \\
  i\frac{da_8}{dt}&=&-\alpha(n_1+1) a_8-\beta a_8-\beta a_6, \nonumber \\
  i\frac{da_9}{dt}&=&-2\beta(n_3+1)a_9.
  \label{coeff_dynamics}
  \end{eqnarray}
The equations for $a_2$ and $a_4$ as well as for $a_6$ and $a_8$ show the
same oscillations, which is obvious as the pair of coefficients $a_2$ and $a_4$ relate states $|1\rangle|2\rangle$
and $|2\rangle|1\rangle$ such that $a_2|1\rangle|2\rangle \Leftrightarrow a_4|2\rangle|1\rangle $ and
the pair of coefficients $a_6$ and $a_8$ relate states $|2\rangle|3\rangle$
and $|3\rangle|2\rangle$  such that $a_6|2\rangle|3\rangle \Leftrightarrow a_8|3\rangle|2\rangle $. The solutions of equations (\ref{coeff_dynamics}) are
\begin{eqnarray}
a_1&=&C_1e^{-2i\alpha n_1t}, \nonumber \\
a_2&=&C_2e^{-i\beta n_3t}+C_4e^{i(2\alpha-\beta n_3)t}, \nonumber \\
a_3&=&C_3e^{-i(\alpha n_1-\beta(n_3+1))t}, \nonumber \\
a_4&=&-C_2e^{-i\beta n_3t}+C_4e^{i(2\alpha-\beta n_3)t}, \nonumber \\
a_5&=&C_5e^{-2i(\alpha (n_1+1)-\beta n_3)t}, \nonumber \\
a_6&=&C_6e^{i\alpha (n_1+1)t}+C_8e^{i(2\beta+\alpha(n_1+1))t}, \nonumber \\
a_7&=&C_3e^{-i(\alpha n_1-\beta(n_3+1))t}, \nonumber \\
a_8&=&-C_6e^{i\alpha (n_1+1)t}+C_8e^{i(2\beta+\alpha(n_1+1))t},\nonumber \\
a_9&=&C_9e^{2i\beta (n_3+1)t}.
\end{eqnarray}
Using these coefficients, we can write the density matrix $\rho$  with elements $\rho_{nm}=a_na_m^*$.
For this density matrix, we can easily calculate the partial transpose matrix with respect to the
spin$_1$. The elements of the partial transpose matrix are connected to the elements $\rho_{nm}$ of the density matrix as
follows:
\begin{eqnarray}
 \langle i_1,j_2|\rho^{T_{\rm spin_1}}|k_1,l_2\rangle\equiv\langle k_1,j_2|\rho|i_1,l_2\rangle.
\end{eqnarray}
The eigenvalues of the matrix $\rho^{T_{\rm spin_1}}$ can be calculated and the negative eigenvalues used in Eq.(\ref{negativity_def}) to calculate the time dependence of negativity.
The value of the Negativity is highly dependent on the choice of initial conditions, i.e., the choice of $C_i, i= 1\cdots9$.

Let us consider the initial state of the system  to be
\begin{eqnarray}
 \psi=a_2(0)|1\rangle|2\rangle+a_6(0)|2\rangle|3\rangle,
\end{eqnarray}
with $a_2(0)=a_6(0)=1/\sqrt{2}$ and
$a_1(0)=a_3(0)=a_4(0)=a_5(0)=a_7(0)=a_8(0)=a_9(0)=0$. For this
choice of initial state, we can calculate the non-zero elements of the density
and the eigenvalues of its partial
transpose with respect to spin$_1$. The partial
transpose matrix has only one negative eigenvalue given as
$\chi=-\sqrt{\rho_{24}\rho_{42}+\rho_{26}\rho_{62}+\rho_{48}\rho_{84}+\rho_{68}\rho_{86}}$
and the negativity from Eq. (\ref{negativity_def}) comes out to be
\begin{figure}[!t]
\includegraphics [angle=0,width=1\columnwidth] {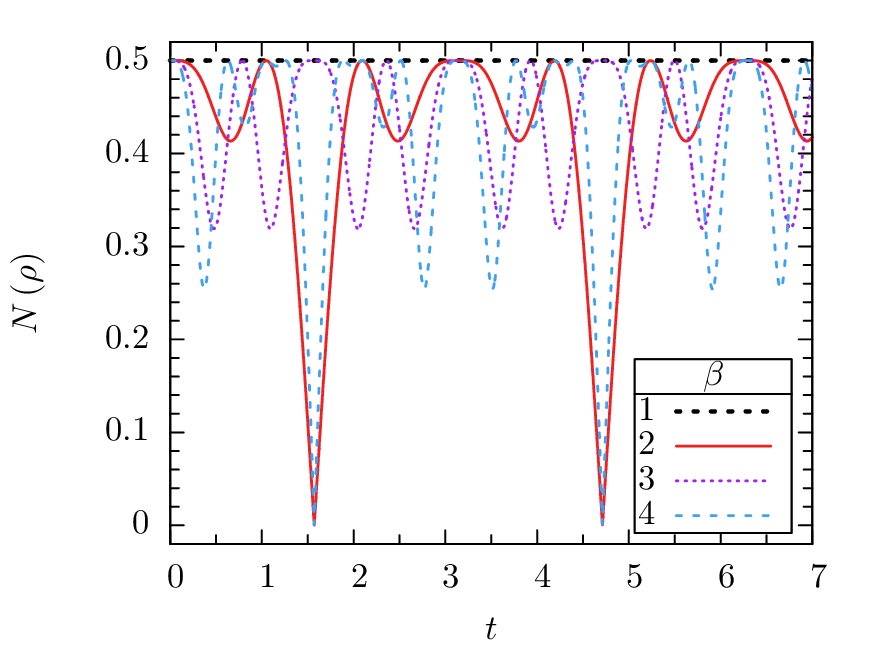}
\caption{(color online) Negativity as a function of time, when
$\alpha=1$ kHz and $\beta= 1, 2, 3, 4$ kHz. The initial conditions are
$a_2(0)=a_6(0)=1/\sqrt{2}$ or $a_4(0)=a_8(0)=1/\sqrt{2}$ and
all the other coefficients are equal to zero. The negativity
vanishes when $\alpha=1$, $\beta=2$ or when $\alpha=1$,
$\beta=4$ at $t=\frac{\pi}{2}+\pi
k,k=0,1,2...$.  The time is measured in milliseconds.}
\label {negativity}
\end{figure}
\begin{eqnarray}
 &&N= \\
&&\frac{\sqrt{6-\cos4\alpha t-\cos4\beta t+2\cos2(\alpha-\beta)t+2\cos2(\alpha+\beta)t}}{4\sqrt{2}}.\nonumber
 \label{negativity2}
\end{eqnarray}
\begin{figure*}[!t]
\centering
  \includegraphics[width=1.0\columnwidth]{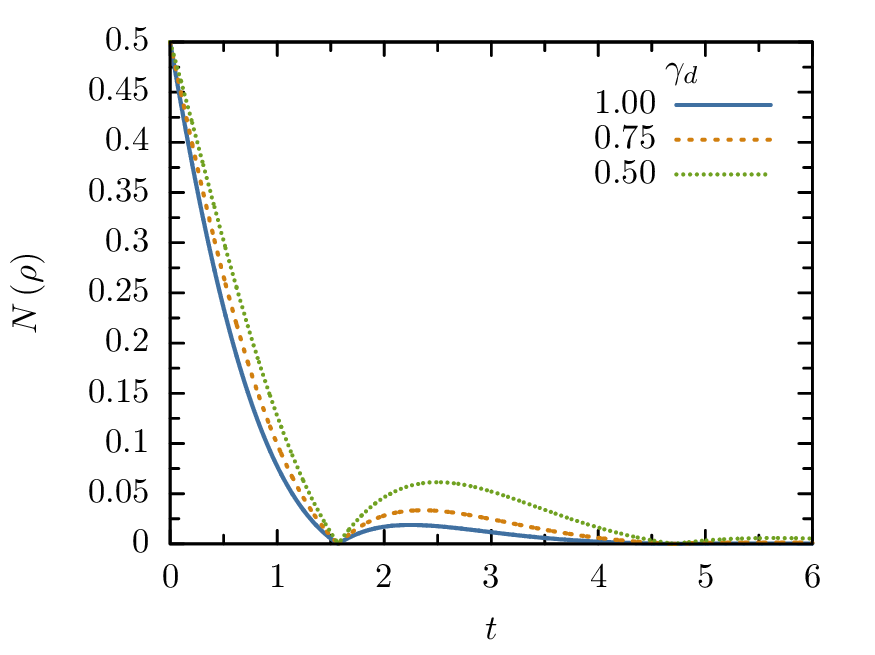} \
    \includegraphics[width=1.0\columnwidth]{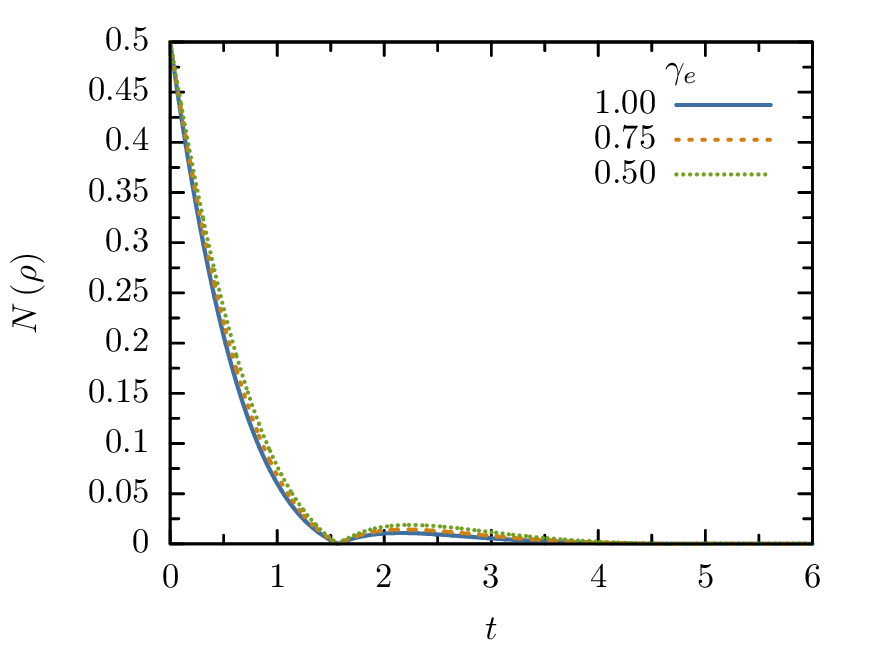}\\
\includegraphics [angle=0,width=1.0\columnwidth]{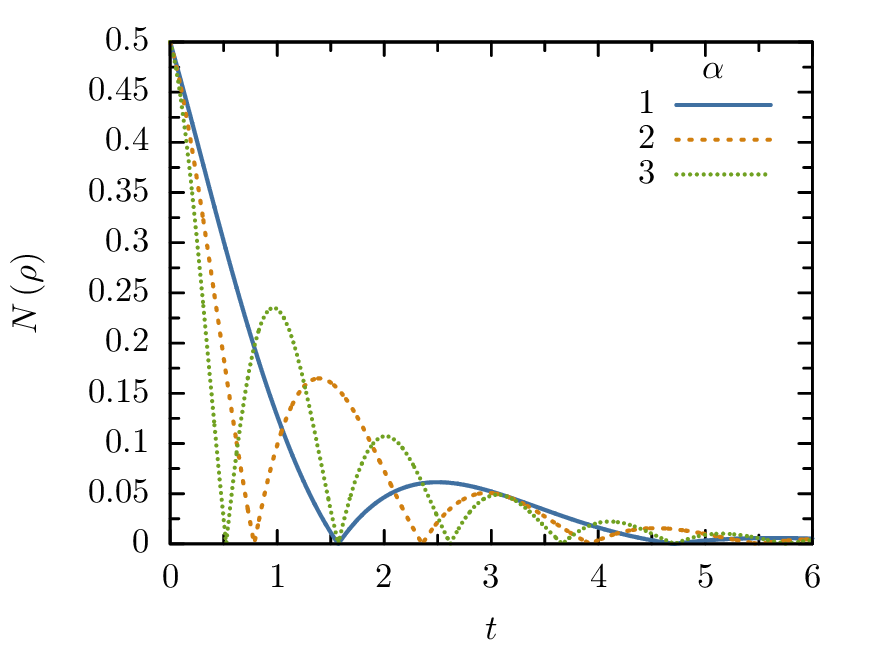}\
\includegraphics [angle=0,width=1.0\columnwidth]{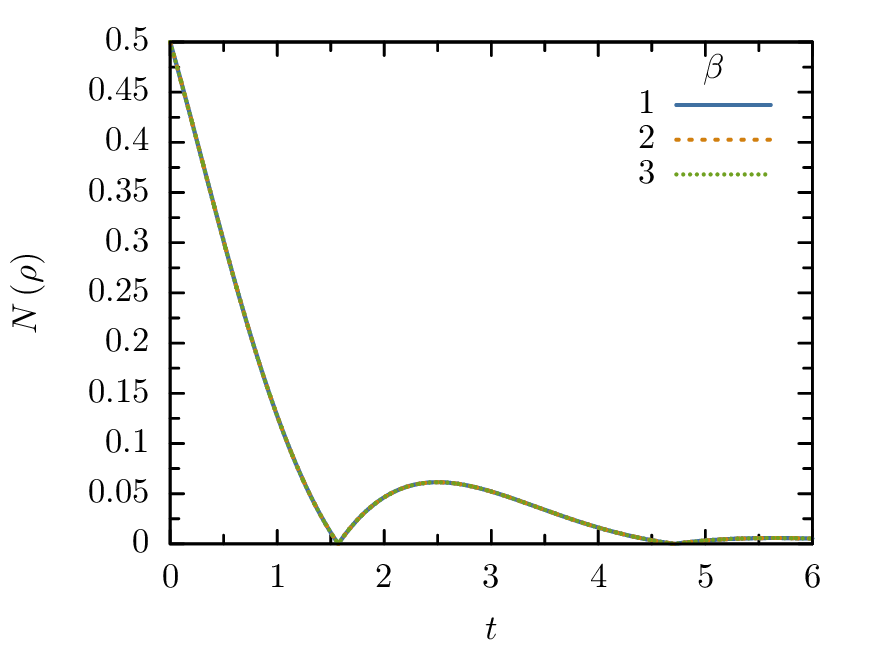}\
  \caption{Time evolution of the negativity of the system in the presence of dissipation is shown. (a)  $\alpha=\beta= \gamma_e=1$ and $\gamma_d$ is varied. (b) $\alpha=\beta= \gamma_d=1$ and $\gamma_e$ is varied. (c) $\beta=1$, $\gamma_e=\gamma_d=0.5$, and $\alpha$ is varied. (d)  $\alpha=1$, $\gamma_e=\gamma_d=0.5$, and $\beta$ is varied.}
      \label{lindblad_sol}
  \end{figure*}
It should be noted that if $\alpha=\beta$ then $N(\rho)=0.5$. We obtain the same result if we take the initial condition
$a_4(0)=a_8(0)=1/\sqrt{2}$ and $a_1(0)=a_2(0)=a_3(0)=a_5(0)=a_6(0)=a_7(0)=a_9(0)=0$.
In Fig. \ref{negativity}, the time dependence of the negativity is shown for a choice of the
initial conditions $a_2(0)=a_6(0)=1/\sqrt{2}$ or  $a_4(0)=a_8(0)=1/\sqrt{2}$ and the rest
of the coefficients being zero. We see that the negativity oscillates in time and the oscillations are more rapid
as the difference between $\alpha$ and $\beta$ increases. In this case, the value of the negativity is bounded between $0$ and $0.5$.

It is interesting to find the instances where the negativity vanishes resulting in zero entanglement between the spins. From Eq.(36) we infer
\begin{eqnarray}
 &&\cos2t(\alpha+\beta)\cos2t(\alpha-\beta)-\cos2t(\alpha+\beta)-\nonumber \\
 &&-\cos2t(\alpha-\beta)=3. \nonumber \\
\end{eqnarray}
It is easy to see that the above equation holds true when
\begin{eqnarray}
 \cos2t(\alpha+\beta)=-1,\nonumber \\
 \cos2t(\alpha-\beta)=-1.
\end{eqnarray}
The solution of these equations delivers the condition for the choice of $\alpha$ and $\beta$ for a zero negativity as
\begin{eqnarray}
 \frac{\alpha}{\beta}=\frac{n+m+1}{n-m},
\end{eqnarray}
where $n$ and $m$ are integers. $n+m+1$ and $n-m$ have different
parity. If the first is even, then the second is odd and vice versa.
Thus, negativity becomes zero when the ratio $\alpha/\beta$
 is equal to a fraction whose denominator is even and the numerator is odd or vice versa.
In Fig.\ref{negativity}, we see that $N(\rho)=0$ when
$\frac{\alpha}{\beta}$ is equal to $\frac{1}{2}$ or $\frac{1}{4}$
and $t=\frac{\pi}{2}+\pi k,k=0,1,2...$

It should be noted here that if we take the initial conditions as $a_2(0)=a_4(0)=a_6(0)=a_8(0)=1/2$ and the rest of the $a_i$ to be zero, the
negativity will be $N(\rho)=1/2$.
For an initial state with all the $a_i$ being the same and equal to $1/3$, the
eigenvalues of the partial transpose of the density matrix cannot be calculated analytically except for the case $\alpha=\beta$. For this case, the negativity
is
\begin{eqnarray}
 N(\rho)=\frac{4}{9}(|\sin\alpha t|-\sin\alpha t),
\end{eqnarray}
which can be expressed in the form
\[
  N(\rho)=\begin{cases}
               0,  \  \  \ \frac{2\pi k}{\alpha}\leq t \leq \frac{(2k+1)\pi}{\alpha}\\
               \frac{8}{9}\sin \alpha t, \  \  \ \frac{(2k+1)\pi}{\alpha} < t < \frac{2(k+1)\pi}{\alpha}
            \end{cases}
\]
where $k$ is an integer. The maximum of the negativity is $N_{\rm max}(\rho)=\frac{8}{9}$ at $t=-\frac{\pi}{2\alpha}+\frac{2\pi k}{\alpha}$ We find that
the entanglement is zero during the time interval $\frac{2\pi k}{\alpha}\leq t \leq \frac{(2k+1)\pi}{\alpha}$, however, there is a partial entanglement sharing
between the spins during the interval $\frac{(2k+1)\pi}{\alpha} < t < \frac{2(k+1)\pi}{\alpha}$.

 \section{Lindblad equation}
 We consider two three-level atoms (NV centers), with the energy levels of each NV center as in Fig.~(3). We have discussed above these two NV centers interacting with each other indirectly via a nanomechanical resonator, a scheme described in
ref. \ref{VEFF}. We now introduce  damping by assuming that the excited levels $|2\rangle$ and $|3\rangle$
decay to the ground state $|1\rangle$, and a direct transition between the excited levels is not allowed. The time evolution of such a system is
given by the master equation (Lindblad Equation) \cite{r38,r39}:
\begin{equation}
\frac{d \rho}{dt}=-i[\hat{V},\rho]+L\rho,
\label{Lindbad}
\end{equation}
where $\hat{V}$ is the (\ref{VEFF}), and $L\rho$ is the damping term\cite{r38,r39}:
\begin{eqnarray}
L\rho=\frac{\gamma_e}{2}\big(2R_{13}^{1}\rho R_{31}^{1}-R_{33}^{1}\rho-\rho R_{33}^{1}\big)+\nonumber \\
+\frac{\gamma_d}{2}\big(2R_{12}^{1}\rho R_{21}^{1}-R_{22}^{1}\rho-\rho R_{22}^{1}\big)+\\
+\frac{\gamma_e}{2}\big(2R_{13}^{2}\rho R_{31}^{2}-R_{33}^{2}\rho-\rho R_{33}^{2}\big)+\nonumber \\
+\frac{\gamma_d}{2}\big(2R_{12}^{2}\rho R_{21}^{2}-R_{22}^{2}\rho-\rho R_{22}^{2}\big)\nonumber .
\end{eqnarray}
The spontaneous emission of atoms 1 and 2 from their excited states $|3\rangle$ to the ground states  $|1\rangle$ is described by the
spontaneous decay rate $\gamma_e$, similarly $\gamma_d$ is the spontaneous
decay rate of excited $|2\rangle$ to the ground $|1\rangle$.
For the derivation of the master equation, we choose the basis:
\begin{equation}
|1\rangle=\Bigg(\begin{array}{lcl}
1\\0\\0\end{array}\Bigg), |2\rangle=\Bigg(\begin{array}{lcl}
0\\1\\0\end{array}\Bigg), |3\rangle=\Bigg(\begin{array}{lcl}
0\\0\\1\end{array}\Bigg)
\label{basis}
\end{equation}
and the $R_{ij}=|i\rangle\langle j|$ operators in this basis have the following forms:
\begin{eqnarray}
&&R_{11}=|1\rangle\langle 1|=\Bigg(\begin{array}{lcl}
1~~0~~0\\0~~0~~0\\0~~0~~0\end{array}\Bigg),R_{22}=|2\rangle\langle 2|=\Bigg(\begin{array}{lcl}
0~~0~~0\\0~~1~~0\\0~~0~~0\end{array}\Bigg),\nonumber \\
&&R_{33}=|3\rangle\langle 3|=\Bigg(\begin{array}{lcl}
0~~0~~0\\0~~0~~0\\0~~0~~1\end{array}\Bigg), R_{13}=|1\rangle\langle 3|=\Bigg(\begin{array}{lcl}
0~~0~~1\\0~~0~~0\\0~~0~~0\end{array}\Bigg), \nonumber \\
&&R_{12}=|1\rangle\langle 2|=\Bigg(\begin{array}{lcl}
0~~1~~0\\0~~0~~0\\0~~0~~0\end{array}\Bigg),R_{21}=|2\rangle\langle 1|=\Bigg(\begin{array}{lcl}
0~~0~~0\\1~~0~~0\\0~~0~~0\end{array}\Bigg),\nonumber \\
&&R_{23}=|2\rangle\langle 3|=\Bigg(\begin{array}{lcl}
0~~0~~0\\0~~0~~1\\0~~0~~0\end{array}\Bigg),R_{32}=|2\rangle\langle 1|=\Bigg(\begin{array}{lcl}
0~~0~~0\\0~~0~~0\\0~~1~~0\end{array}\Bigg),\nonumber \\
&&R_{31}=|3\rangle\langle 1|=\Bigg(\begin{array}{lcl}
0~~0~~0\\0~~0~~0\\1~~0~~0\end{array}\Bigg) \nonumber
\end{eqnarray}
and $R_{ij}^1=R_{ij}\otimes I_{3}$, $R_{ij}^2=I_{3}\otimes R_{ij}$, with $I_{3}$ the three-dimensional unit  matrix.
The most general solution of the master equation (\ref{Lindbad}) for an arbitrary initial state is given in the appendix. In fig. \ref{lindblad_sol}
we have shown the effect of damping on the entanglement between the two spins by varying all the parameters of interest. Fig.~\ref{lindblad_sol}(a) shows
the dynamics of entanglement for different values of $\gamma_d$ with the other parameters fixed as $\alpha=\beta= \gamma_e=1$. We see that the entanglement
decays more sharply due to the increase in decoherence parameter $\gamma_d$. The same can be seen by varying the other decoherence parameter
$\gamma_e$ in fig.~\ref{lindblad_sol}(b). As we see, due to the decoherence, entanglement’s sudden death is smeared out.

 \section{Conclusions}
One of the main challenges for the NV spin-based nanomechanical resonator is to achieve a high degree of entanglement between NV spins. The success of
this proposal naturally depends on the strength of the coupling between NV spins. On the other hand, single NV spins $S=1$ are a triplet state with two
characteristic transition frequencies between the three triplet states. The realization of the controlled transitions between all the three levels requires
two-frequency nanomechanical resonator with a special type of dual frequency cantilever. The frequency characteristics of the cantilever depends on the
particular choice of the cantilever's geometry. A proper setup  of the dual cantilever helps  tuning the  oscillation frequencies to the
resonance frequencies. The dual cantilever is supplemented by the magnetic tips. Thus, the oscillation of the cantilever leads to the indirect interaction
between the NV spins, while the direct interaction is small. Hence, the entanglement between the NV spins is fueled by the dynamics of the cantilever.
In spite of the complexity of the system  (two coupled NV centers and a dual frequency cantilever), the model is analytically  solvable. In order to study decoherence, we utilized the Lindblad master equation. An exact analytical solution of the Lindblad  equation shows the influence of the decoherence.
A prominent effect of the entanglement's sudden death is smeared out by decoherence.

\begin{acknowledgments}
We acknowledge financial support from by Deutsche Forschungsgemeinschaft SFB 762.
\end{acknowledgments}

 \section{Appendix}
 The most general solution of the Lindblad  equation (\ref{Lindbad}) for an
arbitrary initial state is given as ($\rho_{ij}$ on the right-hand
side are the initial values):
 \begin{eqnarray*}
 &&\rho_{99}(t)= \rho_{99}e^{-2\gamma_{e}t};\nonumber \\
 && \rho_{98}(t)=e^{-\frac{\gamma_{d}+3\gamma_{e}}{2}t}(\rho_{98}\cos\beta t-i\rho_{96}\sin\beta t);\nonumber \\
 &&\rho_{97}(t)=\rho_e^{-\frac{3}{2}\gamma_{e}t};\nonumber \\
 && \rho_{96}(t)=e^{-\frac{\gamma_{d}+3\gamma_{e}}{2}t}(\rho_{96}\cos\beta t-i\rho_{98}\sin\beta t);\nonumber \\
 && \rho_{95}(t)=\rho_{95}e^{-(\gamma_{d}+\gamma_{e})t};\nonumber \\
 &&\rho_{94}(t)=e^{-(\frac{\gamma_{d}}{2}+\gamma_{e})t}(\rho_{94}\cos\alpha t-i\rho_{92}\sin\alpha t);\nonumber \\
 && \rho_{93}(t)=\rho_{93}e^{-\frac{3}{2}\gamma_{e}t};\nonumber \\
 &&\rho_{92}(t)=e^{-(\frac{\gamma_{d}}{2}+\gamma_{e})t}(\rho_{92}\cos\alpha t-i\rho_{94}\sin\alpha t);\nonumber \\
 &&\rho_{91}(t)=\rho_{91}e^{-\gamma_{e}t};\nonumber \\
 &&\rho_{89}(t)=e^{-\frac{\gamma_{d}+3\gamma_{e}}{2}t}(\rho_{89}cos\beta t+i \rho_{69}\sin \beta t); \nonumber \\
 &&\rho_{88}(t)=\frac{1}{2}\big(\rho_{66}(1-\cos 2\beta t)+\rho_{88}(1+\cos 2\beta t)+\nonumber \\
 &&+i \rho_{68}\sin 2 \beta t-i\rho_{86}\sin 2 \beta t\big)e^{-(\gamma_{d}+\gamma_{e})t};\nonumber \\
 &&\rho_{87}(t)=e^{-(\frac{\gamma_{d}}{2}+\gamma_{e})t}(\rho_{87}\cos \beta t+i \rho_{67}\sin \beta t);\nonumber \\
 &&\rho_{86}(t)=\frac{1}{2}\big(\rho_{68}(1-\cos 2\beta t)+\rho_{86}(1+\cos 2\beta t)-\nonumber \\
 &&-i \rho_{88}\sin 2 \beta t+i\rho_{66}\sin 2 \beta t\big)e^{-(\gamma_{d}+\gamma_{e})t};\nonumber \\
 &&\rho_{85}(t)=e^{-\frac{3\gamma_{d}+\gamma_{e}}{2}t}(\rho_{85}\cos \beta t+i\rho_{65}\sin \beta t);\nonumber \\
 &&\rho_{84}(t)=e^{-\frac{2\gamma_{d}+\gamma_{e}}{2}t}\big(\sin \alpha t(-i \rho_{82}\cos \beta t+\rho_{62}\sin \beta t)+ \nonumber \\
 &&+\cos \alpha t (\rho_{84}\cos \beta t+i\rho_{64}\sin \beta t)\big);\nonumber \\
 &&\rho_{83}(t)=e^{-\frac{\gamma_{d}+2\gamma_{e}}{2}t}(\rho_{83}\cos \beta t+i \rho_{63} \sin \beta t);\nonumber \\
 &&\rho_{82}(t)=e^{-\frac{2\gamma_{d}+\gamma_{e}}{2}t}\big(\cos \alpha t( \rho_{82}\cos \beta t+i\rho_{62}\sin \beta t)+ \nonumber \\
 &&+\sin \alpha t (-i\rho_{84}\cos \beta t+\rho_{64}\sin \beta t)\big);\nonumber \\
 &&\rho_{81}(t)=e^{-\frac{\gamma_{d}+\gamma_{e}}{2}t}(\rho_{81}\cos \beta t+i \rho_{61}\sin \beta t);\nonumber \\
 &&\rho_{79}(t)=\rho_{79}e^{-\frac{3}{2}\gamma_{e}t};\nonumber \\
 &&\rho_{78}(t)=e^{-\frac{\gamma_{d}+2\gamma_{e}}{2}t}(\rho_{78}\cos \beta t -i\rho_{76}\sin \beta t); \nonumber \\
 &&\rho_{77}(t)=\frac{1}{2}\bigg(2\rho_{77}+\rho_{66}+\rho_{88}+2\rho_{99}-\nonumber\\
 &&-\frac{\gamma_{d}^{2}\big(\rho_{66}-\rho_{88}\big)}{4\beta^2+\gamma_d^2}-\frac{2i\beta\gamma_{d}\big(\rho_{86}-\rho_{68}\big)}{4\beta^2+\gamma_d^2}\bigg)
 e^{-\gamma_{e}t}- \nonumber \\
 &&-\frac{\gamma_d}{2}\big(\rho_{66}-\rho_{88}\big)\frac{2\beta\sin 2\beta t-\gamma_{d}\cos2 \beta t}{4\beta^2+\gamma_d^2}e^{-(\gamma_d+\gamma_e)t}+\nonumber \\
 &&+\frac{i\gamma_d}{2}\big(\rho_{86}-\rho_{68}\big)\frac{2\beta\cos 2\beta t+\gamma_{d}\sin2 \beta t}{4\beta^2+\gamma_d^2}e^{-(\gamma_d+\gamma_e)t}-\nonumber\\
 &&-\frac{1}{2}\big(\rho_{66}+\rho_{88}\big)e^{-(\gamma_e+\gamma_d)t}-\rho_{99}e^{-2\gamma_e t};\nonumber\\
  \end{eqnarray*}
 \begin{eqnarray}
 &&\rho_{76}(t)=e^{-\frac{\gamma_{d}+2\gamma_{e}}{2}t}(\rho_{76}\cos \beta t -i\rho_{78}\sin \beta t); \nonumber \\
  &&\rho_{75}(t)=\rho_{75}e^{-\frac{2\gamma_{d}+\gamma_{e}}{2}t};\nonumber \\
  &&\rho_{74}(t)=e^{-\frac{\gamma_d+\gamma_e}{2}t}\big(\gamma_{74}\cos \alpha t-i\gamma_{72}\sin \alpha t)+\nonumber\\
  &&+e^{-\frac{\gamma_d+\gamma_e}{2}t}\Bigg(\cos \alpha t\int e^{\frac{\gamma_d+\gamma_e}{2}t}F_{74}(t)\cos \alpha t+\nonumber \\
  &&+\sin \alpha t\int e^{\frac{\gamma_d+\gamma_e}{2}t}F_{74}(t)\sin \alpha t\Bigg);\nonumber \\
  &&\rho_{73}(t)=\rho_{73}e^{-\gamma_e t};\nonumber \\
  &&\rho_{72}(t)=e^{-\frac{\gamma_d+\gamma_e}{2}t}\big(\gamma_{72}\cos \alpha t-i\gamma_{74}\sin \alpha t)+\nonumber\\
  &&+ie^{-\frac{\gamma_d+\gamma_e}{2}t}\Bigg(\cos \alpha t\int e^{\frac{\gamma_d+\gamma_e}{2}t}F_{74}(t)\sin \alpha t-\nonumber \\
  &&-\sin \alpha t\int e^{\frac{\gamma_d+\gamma_e}{2}t}F_{74}(t)\cos \alpha t\Bigg);\nonumber \\
  &&F_{74}(t)=\gamma_d \rho_{85}(t)+\gamma_e \rho_{96}(t);\nonumber\\
  &&2\int e^{\frac{\gamma_d+\gamma_e}{2}t}F_{74}(t)\sin \alpha t=\nonumber\\
  &&=-\gamma_{d}\rho_{85}e^{-\gamma_d t}\Big(\frac{(\alpha-\beta)\cos(\alpha-\beta)t+\gamma_d \sin(\alpha-\beta)t}{(\alpha-\beta)^2+\gamma_d^2}+\nonumber\\
  &&+\frac{(\alpha+\beta)\cos(\alpha+\beta)t+\gamma_d \sin(\alpha+\beta)t}{(\alpha+\beta)^2+\gamma_d^2}\Big)+\nonumber \\
  &&+i\gamma_{d}\rho_{65}e^{-\gamma_d t}\Big(\frac{(\alpha-\beta)\sin(\alpha-\beta)t-\gamma_d \cos(\alpha-\beta)t}{(\alpha-\beta)^2+\gamma_d^2}-\nonumber\\
  &&-\frac{(\alpha+\beta)\sin(\alpha+\beta)t-\gamma_d \cos(\alpha+\beta)t}{(\alpha+\beta)^2+\gamma_d^2}\Big)-\nonumber \\
  &&-\gamma_{e}\rho_{96}e^{-\gamma_e t}\Big(\frac{(\alpha-\beta)\cos(\alpha-\beta)t+\gamma_e \sin(\alpha-\beta)t}{(\alpha-\beta)^2+\gamma_e^2}+\nonumber\\
  &&+\frac{(\alpha+\beta)\cos(\alpha+\beta)t+\gamma_e \sin(\alpha+\beta)t}{(\alpha+\beta)^2+\gamma_e^2}\Big)-\nonumber \\
  &&-i\gamma_{e}\rho_{98}e^{-\gamma_e t}\Big(\frac{(\alpha-\beta)\sin(\alpha-\beta)t-\gamma_e \cos(\alpha-\beta)t}{(\alpha-\beta)^2+\gamma_e^2}-\nonumber\\
  &&-\frac{(\alpha+\beta)\sin(\alpha+\beta)t-\gamma_e \cos(\alpha+\beta)t}{(\alpha+\beta)^2+\gamma_e^2}\Big);\nonumber \\
  &&2\int e^{\frac{\gamma_d+\gamma_e}{2}t}F_{74}(t)\cos \alpha t=\nonumber\\
  &&=\gamma_{d}\rho_{85}e^{-\gamma_d t}\Big(\frac{(\alpha-\beta)\sin(\alpha-\beta)t-\gamma_d \cos(\alpha-\beta)t}{(\alpha-\beta)^2+\gamma_d^2}+\nonumber\\
  &&+\frac{(\alpha+\beta)\sin(\alpha+\beta)t-\gamma_d \cos(\alpha+\beta)t}{(\alpha+\beta)^2+\gamma_d^2}\Big)+\nonumber \\
  &&+i\gamma_{d}\rho_{65}e^{-\gamma_d t}\Big(\frac{(\alpha-\beta)\cos(\alpha-\beta)t+\gamma_d \sin(\alpha-\beta)t}{(\alpha-\beta)^2+\gamma_d^2}-\nonumber\\
  &&-\frac{(\alpha+\beta)\cos(\alpha+\beta)t+\gamma_d \sin(\alpha+\beta)t}{(\alpha+\beta)^2+\gamma_d^2}\Big)+\nonumber \\
  &&+\gamma_{e}\rho_{96}e^{-\gamma_e t}\Big(\frac{(\alpha-\beta)\sin(\alpha-\beta)t-\gamma_e \cos(\alpha-\beta)t}{(\alpha-\beta)^2+\gamma_e^2}+\nonumber\\
  &&+\frac{(\alpha+\beta)\sin(\alpha+\beta)t-\gamma_e \cos(\alpha+\beta)t}{(\alpha+\beta)^2+\gamma_e^2}\Big)-\nonumber \\
  &&-i\gamma_{e}\rho_{98}e^{-\gamma_e t}\Big(\frac{(\alpha-\beta)\cos(\alpha-\beta)t+\gamma_e \sin(\alpha-\beta)t}{(\alpha-\beta)^2+\gamma_e^2}-\nonumber\\
  &&-\frac{(\alpha+\beta)\cos(\alpha+\beta)t+\gamma_e \sin(\alpha+\beta)t}{(\alpha+\beta)^2+\gamma_e^2}\Big);\nonumber
 \end{eqnarray}
 \begin{eqnarray}
  &&\gamma_{72}=\rho_{72}-\frac{1}{2}\gamma_d\rho_{65}\Bigg(\frac{\gamma_d}{(\alpha+\beta)^2+\gamma_d^2}+\frac{\gamma_d}{(\alpha-\beta)^2+\gamma_d^2}\Bigg)+\nonumber\\
   &&+\frac{1}{2}\gamma_e\rho_{98}\Bigg(\frac{\gamma_e}{(\alpha+\beta)^2+\gamma_d^2}+\frac{\gamma_e}{(\alpha-\beta)^2+\gamma_d^2}\Bigg)+\nonumber\\
  &&+\frac{i}{2}\gamma_{d}\rho_{85}\Bigg(\frac{\alpha-\beta}{(\alpha-\beta)^2+\gamma_d^2}-\frac{\alpha+\beta}{(\alpha+\beta)^2+\gamma_d^2}\Bigg)+\nonumber\\
  &&+\frac{i}{2}\gamma_{e}\rho_{96}\Bigg(\frac{\alpha-\beta}{(\alpha-\beta)^2+\gamma_e^2}-\frac{\alpha+\beta}{(\alpha+\beta)^2+\gamma_e^2}\Bigg); \nonumber \\
   &&\gamma_{74}=\rho_{74}+\frac{1}{2}\gamma_d\rho_{85}\Bigg(\frac{\gamma_d}{(\alpha+\beta)^2+\gamma_d^2}+\frac{\gamma_d}{(\alpha-\beta)^2+\gamma_d^2}\Bigg)+\nonumber\\
   &&+\frac{1}{2}\gamma_e\rho_{96}\Bigg(\frac{\gamma_e}{(\alpha+\beta)^2+\gamma_d^2}+\frac{\gamma_e}{(\alpha-\beta)^2+\gamma_d^2}\Bigg)-\nonumber\\
  &&-\frac{i}{2}\gamma_{d}\rho_{65}\Bigg(\frac{\alpha-\beta}{(\alpha-\beta)^2+\gamma_d^2}+\frac{\alpha+\beta}{(\alpha+\beta)^2+\gamma_d^2}\Bigg)+\nonumber\\
  &&+\frac{i}{2}\gamma_{e}\rho_{98}\Bigg(\frac{\alpha-\beta}{(\alpha-\beta)^2+\gamma_e^2}+\frac{\alpha+\beta}{(\alpha+\beta)^2+\gamma_e^2}\Bigg); \nonumber
  \end{eqnarray}
 \begin{eqnarray}
   &&\rho_{71}(t)=\gamma_{71}e^{-\frac{\gamma_e}{2}t}-\rho_{93}e^{-\frac{3\gamma_e t}{2}}+\frac{\gamma_d}{2}e^{-\frac{2\gamma_d+\gamma_e}{2}t}\nonumber\\
   &&\Bigg(\rho_{82}\Big(\frac{(\alpha-\beta)\sin(\alpha-\beta)t-\gamma_d\cos(\alpha-\beta)t}{(\alpha-\beta)^2+\gamma_d^2}+\nonumber\\
   &&+\frac{(\alpha+\beta)\sin(\alpha+\beta)t-\gamma_d\cos(\alpha+\beta)t}{(\alpha+\beta)^2+\gamma_d^2}\Big)+\nonumber \\
   &&+i\rho_{62}\Big(\frac{(\alpha-\beta)\cos(\alpha-\beta)t+\gamma_d\sin(\alpha-\beta)t}{(\alpha-\beta)^2+\gamma_d^2}-\nonumber\\
   &&-\frac{(\alpha+\beta)\cos(\alpha+\beta)t+\gamma_d\cos(\alpha+\beta)t}{(\alpha+\beta)^2+\gamma_d^2}\Big)+\nonumber \\
   &&+i\rho_{84}\Big(\frac{(\alpha-\beta)\cos(\alpha-\beta)t+\gamma_d\sin(\alpha-\beta)t}{(\alpha-\beta)^2+\gamma_d^2}+\nonumber\\
   &&+\frac{(\alpha+\beta)\cos(\alpha+\beta)t+\gamma_d\cos(\alpha+\beta)t}{(\alpha+\beta)^2+\gamma_d^2}\Big)+\nonumber \\
   &&\rho_{64}\Big(\frac{(\alpha-\beta)\sin(\alpha-\beta)t-\gamma_d\cos(\alpha-\beta)t}{(\alpha-\beta)^2+\gamma_d^2}-\nonumber\\
   &&-\frac{(\alpha+\beta)\sin(\alpha+\beta)t-\gamma_d\cos(\alpha+\beta)t}{(\alpha+\beta)^2+\gamma_d^2}\Big)\Bigg);\nonumber \\
   &&\gamma_{71}=\rho_{71}+\rho_{93}+\nonumber\\
   &&+\frac{\gamma_d}{2}\Bigg(\rho_{82}\bigg(\frac{\gamma_d}{(\alpha-\beta)^2+\gamma_d^2}+\frac{\gamma_d}{(\alpha+\beta)^2+\gamma_d^2}\bigg)-\nonumber\\
   &&-i\rho_{62}\bigg(\frac{\alpha-\beta}{(\alpha-\beta)^2+\gamma_d^2}-\frac{\alpha+\beta}{(\alpha+\beta)^2+\gamma_d^2}\bigg)-\nonumber\\
   &&-i\rho_{84}\bigg(\frac{\alpha-\beta}{(\alpha-\beta)^2+\gamma_d^2}+\frac{\alpha+\beta}{(\alpha+\beta)^2+\gamma_d^2}\bigg)+\nonumber\\
   &&+\rho_{64}\bigg(\frac{\gamma_d}{(\alpha-\beta)^2+\gamma_d^2}-\frac{\gamma_d}{(\alpha+\beta)^2+\gamma_d^2}\bigg)\Bigg);\nonumber
   \end{eqnarray}
 \begin{eqnarray}
 &&\rho_{69}(t)=e^{-\frac{\gamma_{d}+3\gamma_{e}}{2}t}(\rho_{69}\cos\beta t+i \rho_{89}\sin \beta t); \nonumber \\
 &&\rho_{68}(t)=\frac{1}{2}\big(\rho_{86}(1-\cos 2 \beta t)+\rho_{68}(1+\cos 2\beta t)+\nonumber \\
 &&+i \rho_{88}\sin 2 \beta t-i \rho_{66} \sin 2 \beta t\big)e^{-(\gamma_{d}+\gamma_{e})t}; \nonumber \\
 &&\rho_{67}(t)=e^{-(\frac{\gamma_{d}}{2}+\gamma_{e})t}(\rho_{67} \cos \beta t+i \rho_{87}\sin \beta t);\nonumber\\
 &&\rho_{66}(t)=\frac{1}{2}\big(\rho_{88}(1-\cos 2 \beta t)+\rho_{66}(1+\cos 2\beta t)-\nonumber \\
 &&-i \rho_{68}\sin 2 \beta t+i \rho_{86} \sin 2 \beta t\big)e^{-(\gamma_{d}+\gamma_{e})t}; \nonumber \\
 &&\rho_{65}(t)=e^{-\frac{3\gamma_{d}+\gamma_{e}}{2}t}(\rho_{65}\cos \beta t+i\rho_{85}\sin \beta t);\nonumber \\
 &&\rho_{64}(t)=e^{-\frac{2\gamma_{d}+\gamma_{e}}{2}t}\big(\cos \alpha t( \rho_{64}\cos \beta t+i\rho_{84}\sin \beta t)+ \nonumber \\
 &&+\sin \alpha t (-i\rho_{62}\cos \beta t+\rho_{82}\sin \beta t)\big);\nonumber \\
 &&\rho_{63}(t)=e^{-\frac{\gamma_{d}+2\gamma_{e}}{2}t}(\rho_{63}\cos \beta t+i \rho_{83} \sin \beta t);\nonumber \\
 &&\rho_{62}(t)=e^{-\frac{2\gamma_{d}+\gamma_{e}}{2}t}\big(\sin \alpha t(-i \rho_{64}\cos \beta t+\rho_{84}\sin \beta t)+ \nonumber \\
 &&+\cos \alpha t (\rho_{62}\cos \beta t+i\rho_{82}\sin \beta t)\big);\nonumber \\
 &&\rho_{61}(t)=e^{-\frac{\gamma_{d}+\gamma_{e}}{2}t}(\rho_{61}\cos \beta t+i \rho_{81}\sin \beta t);\nonumber\\
 &&\rho_{59}(t)=\rho_{59}e^{-(\gamma_{d}+\gamma_{e})t};\nonumber \\
 &&\rho_{58}(t)=e^{-\frac{3\gamma_{d}+\gamma_{e}}{2}t}(\rho_{58}\cos\beta t-i \rho_{56}\sin \beta t); \nonumber \\
  &&\rho_{57}(t)=\rho_{57}e^{-\frac{2\gamma_{d}+\gamma_{e}}{2}t};\nonumber \\
  &&\rho_{56}(t)=e^{-\frac{3\gamma_{d}+\gamma_{e}}{2}t}(\rho_{56}\cos\beta t-i \rho_{58}\sin \beta t);\nonumber \\
  &&\rho_{55}(t)=\rho_{55}e^{-2\gamma_{d}t}; \nonumber \\
  &&\rho_{54}(t)=e^{-\frac{3\gamma_{d}}{2}t}(\rho_{54}\cos \alpha t-i \rho_{52}\sin \alpha t); \nonumber \\
  &&\rho_{53}(t)=\rho_{53}e^{-\frac{2\gamma_{d}+\gamma_{e}}{2}t};\nonumber \\
  &&\rho_{52}(t)=e^{-\frac{3\gamma_{d}}{2}t}(\rho_{52}\cos \alpha t-i \rho_{54}\sin \alpha t); \nonumber \\
  &&\rho_{51}(t)=\rho_{51}e^{-\gamma_{d}t};\nonumber \\
  &&\rho_{49}(t)=e^{-\frac{\gamma_{d}+2\gamma_{e}}{2}t}(\rho_{49}\cos \alpha t-i\rho_{29} \sin \alpha t);\nonumber \\
   &&\rho_{48}(t)=e^{-\frac{2\gamma_{d}+\gamma_{e}}{2}t}\big(\sin \alpha t(i \rho_{28}\cos \beta t+\rho_{26}\rho_{28}\sin \beta t)+\nonumber\\
   &&+\cos \alpha t(\rho_{48}\cos \beta t-i \rho_{46}\sin \beta t)\big);\nonumber\\
     &&\rho_{47}(t)=e^{-\frac{\gamma_d+\gamma_e}{2}t}\big(\gamma_{47}\cos \alpha t+i \gamma_{27}\sin \alpha t\big)+\nonumber\\
  &&+e^{-\frac{\gamma_d+\gamma_e}{2}t}\big(\cos \alpha t\int e^{\frac{\gamma_d+\gamma_e}{2}t}F_{47}(t)\cos \alpha t+\nonumber \\
    &&+\sin \alpha t\int e^{\frac{\gamma_d+\gamma_e}{2}t}F_{47}(t)\sin \alpha t\big);\nonumber \\
    &&F_{47}(t)=\gamma_{d}\rho_{58}(t)+\gamma_{e}\rho_{69}(t);\nonumber \\
    &&\rho_{46}(t)=e^{-\frac{2\gamma_{d}+\gamma_{e}}{2}t}\big(\cos\alpha t(\rho_{46}\cos \beta t-i \rho_{48} \sin \beta t)+\nonumber\\
   &&+\sin\alpha t(i\rho_{26}\cos \beta t+\rho_{28}\sin \beta t)\big);\nonumber \\
    &&\rho_{45}(t)=e^{-\frac{3}{2}\gamma_d t}(\rho_{45}\cos \alpha t+i\rho_{25}\sin \alpha t);\nonumber
  \end{eqnarray}
 \begin{eqnarray}
 &&2\int e^{\frac{\gamma_d+\gamma_e}{2}t}F_{47}(t)\sin \alpha t=\nonumber\\
  &&=-\gamma_{d}\rho_{58}e^{-\gamma_d t}\Big(\frac{(\alpha-\beta)\cos(\alpha-\beta)t+\gamma_d \sin(\alpha-\beta)t}{(\alpha-\beta)^2+\gamma_d^2}+\nonumber\\
  &&+\frac{(\alpha+\beta)\cos(\alpha+\beta)t+\gamma_d \sin(\alpha+\beta)t}{(\alpha+\beta)^2+\gamma_d^2}\Big)-\nonumber \\
  &&-i\gamma_{d}\rho_{56}e^{-\gamma_d t}\Big(\frac{(\alpha-\beta)\sin(\alpha-\beta)t-\gamma_d \cos(\alpha-\beta)t}{(\alpha-\beta)^2+\gamma_d^2}-\nonumber\\
  &&-\frac{(\alpha+\beta)\sin(\alpha+\beta)t-\gamma_d \cos(\alpha+\beta)t}{(\alpha+\beta)^2+\gamma_d^2}\Big)-\nonumber \\
  &&-\gamma_{e}\rho_{69}e^{-\gamma_e t}\Big(\frac{(\alpha-\beta)\cos(\alpha-\beta)t+\gamma_e \sin(\alpha-\beta)t}{(\alpha-\beta)^2+\gamma_e^2}+\nonumber\\
  &&+\frac{(\alpha+\beta)\cos(\alpha+\beta)t+\gamma_e \sin(\alpha+\beta)t}{(\alpha+\beta)^2+\gamma_e^2}\Big)+\nonumber \\
  &&+i\gamma_{e}\rho_{89}e^{-\gamma_e t}\Big(\frac{(\alpha-\beta)\sin(\alpha-\beta)t-\gamma_e \cos(\alpha-\beta)t}{(\alpha-\beta)^2+\gamma_e^2}-\nonumber\\
  &&-\frac{(\alpha+\beta)\sin(\alpha+\beta)t-\gamma_e \cos(\alpha+\beta)t}{(\alpha+\beta)^2+\gamma_e^2}\Big);\nonumber \\
    &&2\int e^{\frac{\gamma_d+\gamma_e}{2}t}F_{47}(t)\cos \alpha t=\nonumber\\
  &&=\gamma_{d}\rho_{58}e^{-\gamma_d t}\Big(\frac{(\alpha-\beta)\sin(\alpha-\beta)t-\gamma_d \cos(\alpha-\beta)t}{(\alpha-\beta)^2+\gamma_d^2}+\nonumber\\
  &&+\frac{(\alpha+\beta)\sin(\alpha+\beta)t-\gamma_d \cos(\alpha+\beta)t}{(\alpha+\beta)^2+\gamma_d^2}\Big)-\nonumber \\
  &&-i\gamma_{d}\rho_{56}e^{-\gamma_d t}\Big(\frac{(\alpha-\beta)\cos(\alpha-\beta)t+\gamma_d \sin(\alpha-\beta)t}{(\alpha-\beta)^2+\gamma_d^2}-\nonumber\\
  &&-\frac{(\alpha+\beta)\cos(\alpha+\beta)t+\gamma_d \sin(\alpha+\beta)t}{(\alpha+\beta)^2+\gamma_d^2}\Big)+\nonumber \\
  &&+\gamma_{e}\rho_{69}e^{-\gamma_e t}\Big(\frac{(\alpha-\beta)\sin(\alpha-\beta)t-\gamma_e \cos(\alpha-\beta)t}{(\alpha-\beta)^2+\gamma_e^2}+\nonumber\\
  &&+\frac{(\alpha+\beta)\sin(\alpha+\beta)t-\gamma_e \cos(\alpha+\beta)t}{(\alpha+\beta)^2+\gamma_e^2}\Big)+\nonumber \\
  &&+i\gamma_{e}\rho_{89}e^{-\gamma_e t}\Big(\frac{(\alpha-\beta)\cos(\alpha-\beta)t+\gamma_e \sin(\alpha-\beta)t}{(\alpha-\beta)^2+\gamma_e^2}-\nonumber\\
  &&-\frac{(\alpha+\beta)\cos(\alpha+\beta)t+\gamma_e \sin(\alpha+\beta)t}{(\alpha+\beta)^2+\gamma_e^2}\Big);\nonumber\\
   &&\gamma_{27}=\rho_{27}+\frac{1}{2}\gamma_d\rho_{56}\Bigg(\frac{\gamma_d}{(\alpha+\beta)^2+\gamma_d^2}-\frac{\gamma_d}{(\alpha-\beta)^2+\gamma_d^2}\Bigg)-\nonumber\\
   &&-\frac{1}{2}\gamma_e\rho_{89}\Bigg(\frac{\gamma_e}{(\alpha+\beta)^2+\gamma_d^2}-\frac{\gamma_e}{(\alpha-\beta)^2+\gamma_d^2}\Bigg)-\nonumber\\
  &&-\frac{i}{2}\gamma_{d}\rho_{58}\Bigg(\frac{\alpha-\beta}{(\alpha-\beta)^2+\gamma_d^2}+\frac{\alpha+\beta}{(\alpha+\beta)^2+\gamma_d^2}\Bigg)-\nonumber\\
  &&-\frac{i}{2}\gamma_{e}\rho_{69}\Bigg(\frac{\alpha-\beta}{(\alpha-\beta)^2+\gamma_e^2}+\frac{\alpha+\beta}{(\alpha+\beta)^2+\gamma_e^2}\Bigg); \nonumber
 \end{eqnarray}
 \newpage
 \begin{eqnarray}
   &&\gamma_{47}=\rho_{47}+\frac{1}{2}\gamma_d\rho_{58}\Bigg(\frac{\gamma_d}{(\alpha+\beta)^2+\gamma_d^2}+\frac{\gamma_d}{(\alpha-\beta)^2+\gamma_d^2}\Bigg)+\nonumber\\
   &&+\frac{1}{2}\gamma_e\rho_{69}\Bigg(\frac{\gamma_e}{(\alpha+\beta)^2+\gamma_d^2}+\frac{\gamma_e}{(\alpha-\beta)^2+\gamma_d^2}\Bigg)+\nonumber\\
  &&+\frac{i}{2}\gamma_{d}\rho_{56}\Bigg(\frac{\alpha-\beta}{(\alpha-\beta)^2+\gamma_d^2}-\frac{\alpha+\beta}{(\alpha+\beta)^2+\gamma_d^2}\Bigg)+\nonumber\\
  &&-\frac{i}{2}\gamma_{e}\rho_{89}\Bigg(\frac{\alpha-\beta}{(\alpha-\beta)^2+\gamma_e^2}-\frac{\alpha+\beta}{(\alpha+\beta)^2+\gamma_e^2}\Bigg);
  \nonumber \\
 &&\rho_{44}(t)=\frac{1}{2}\big(\rho_{22}+\rho_{44}+2\rho_{55}+\rho_{88}+\rho_{66}\big)e^{-\gamma_d t}-\nonumber \\
 &&-\rho_{55}e^{-2\gamma_d t}-\frac{1}{2}(\rho_{88}+\rho_{66})e^{-(\gamma_d+\gamma_e) t}-\nonumber \\
 &&-\frac{e^{-\gamma_d t}}{2}\big(\gamma_x \cos 2\alpha t-i\gamma_y \sin 2\alpha t\big)-\nonumber \\
 &&-\frac{\gamma_d e^{-\gamma_d t}}{2}\big(\cos 2\alpha t \int e^{\gamma_d t}(\rho_{88}(t)-\rho_{66}(t))\cos 2\alpha t dt+\nonumber \\
 &&+\sin 2\alpha t \int e^{\gamma_d t}(\rho_{88}(t)-\rho_{66}(t))\sin 2\alpha t dt\big);\nonumber \\
 && \int e^{\gamma_d t}(\rho_{88}(t)-\rho_{66}(t))\sin 2\alpha t dt=\frac{(\rho_{88}-\rho_{66})}{2}e^{-\gamma_e t}\nonumber \\
 && \bigg(-\frac{2(\alpha-\beta)\cos2t(\alpha-\beta)+\gamma_e \sin 2t(\alpha-\beta)}{4(\alpha-\beta)^2+\gamma_e^2}-\nonumber \\
 &&-\frac{2(\alpha+\beta)\cos2t(\alpha+\beta)+\gamma_e \sin 2t(\alpha+\beta)}{4(\alpha+\beta)^2+\gamma_e^2}\bigg)+\nonumber \\
 &&+i\frac{(\rho_{68}-\rho_{86})}{2}e^{-\gamma_e t}\nonumber \\
 && \bigg(\frac{2(\alpha-\beta)\sin2t(\alpha-\beta)-\gamma_e \cos 2t(\alpha-\beta)}{4(\alpha-\beta)^2+\gamma_e^2}-\nonumber \\
 &&-\frac{2(\alpha+\beta)\sin2t(\alpha+\beta)-\gamma_e \cos 2t(\alpha+\beta)}{4(\alpha+\beta)^2+\gamma_e^2}\bigg); \nonumber \\
 && \int e^{\gamma_d t}(\rho_{88}(t)-\rho_{66}(t))\cos 2\alpha t dt=\frac{(\rho_{88}-\rho_{66})}{2}e^{-\gamma_e t}\nonumber \\
 && \bigg(\frac{2(\alpha-\beta)\sin2t(\alpha-\beta)-\gamma_e \cos 2t(\alpha-\beta)}{4(\alpha-\beta)^2+\gamma_e^2}+\nonumber \\
 &&+\frac{2(\alpha+\beta)\sin2t(\alpha+\beta)-\gamma_e \cos 2t(\alpha+\beta)}{4(\alpha+\beta)^2+\gamma_e^2}\bigg)+\nonumber \\
 &&+i\frac{(\rho_{68}-\rho_{86})}{2}e^{-\gamma_e t}\nonumber \\
 && \bigg(\frac{2(\alpha-\beta)\cos2t(\alpha-\beta)+\gamma_e \sin 2t(\alpha-\beta)}{4(\alpha-\beta)^2+\gamma_e^2}-\nonumber \\
 &&-\frac{2(\alpha+\beta)\cos2t(\alpha+\beta)+\gamma_e \sin 2t(\alpha+\beta)}{4(\alpha+\beta)^2+\gamma_e^2}\bigg);\nonumber \\
 && \gamma_x =\rho_{22}-\rho_{44}+\gamma_d\frac{\rho_{88}-\rho_{66}}{2}\nonumber \\
 &&\Bigg(\frac{\gamma_e}{4(\alpha+\beta)^2+\gamma_e^2}+\frac{\gamma_e}{4(\alpha-\beta)^2+\gamma_e^2}\Bigg)-\nonumber \\
 &&-i \gamma_d\frac{\rho_{68}-\rho_{86}}{2}\nonumber \\
 && \Bigg(\frac{2(\alpha-\beta)}{4(\alpha-\beta)^2+\gamma_e^2}-\frac{2(\alpha+\beta)}{4(\alpha+\beta)^2+\gamma_e^2}\Bigg);\nonumber
 \end{eqnarray}
 \begin{eqnarray}
 && \gamma_y =\rho_{24}-\rho_{42}+i\gamma_d\frac{\rho_{88}-\rho_{66}}{2}\nonumber \\
 &&\Bigg(\frac{2(\alpha-\beta)}{4(\alpha-\beta)^2+\gamma_e^2}+\frac{2(\alpha+\beta)}{4(\alpha+\beta)^2+\gamma_e^2}\Bigg)-\nonumber \\
 &&- \gamma_d\frac{\rho_{68}-\rho_{86}}{2}\nonumber \\
 &&\Bigg(\frac{\gamma_e}{4(\alpha-\beta)^2+\gamma_e^2}-\frac{\gamma_e}{4(\alpha+\beta)^2+\gamma_e^2}\Bigg);\nonumber\\
&&\rho_{43}(t)=e^{-\frac{\gamma_d+\gamma_e}{2}t}\Bigg\{\gamma_{43}\cos \alpha t+i\gamma_{23}\sin \alpha t-\nonumber \\
&&-i\cos \alpha t \int e^{\frac{\gamma_d+\gamma_e}{2}t}F_{23}(t)\sin \alpha t dt+\nonumber\\
&&+i\sin \alpha t \int e^{\frac{\gamma_d+\gamma_e}{2}t}F_{23}(t)\cos \alpha t dt\Bigg\};\nonumber \\
&&F_{23}(t)=\gamma_d \rho_{56}(t)+\gamma_e \rho_{89}(t);\nonumber \\
&&\int e^{\frac{\gamma_d+\gamma_e}{2}t}F_{23}(t)\cos \alpha t dt=\nonumber \\
&&=\frac{\gamma_d \rho_{56}e^{-\gamma_d t}}{2}\Bigg(\frac{(\alpha-\beta)\sin(\alpha-\beta)t-\gamma_d\cos(\alpha-\beta)t}{(\alpha-\beta)^2+\gamma_d^2}+\nonumber \\
&&+\frac{(\alpha+\beta)\sin(\alpha+\beta)t-\gamma_d\cos(\alpha+\beta)t}{(\alpha+\beta)^2+\gamma_d^2}\bigg)-\nonumber \\
&&-\frac{i \gamma_d\rho_{58}e^{-\gamma_dt}}{2}\Bigg(\frac{(\alpha-\beta)\cos(\alpha-\beta)t+\gamma_d\sin(\alpha-\beta)t}{(\alpha-\beta)^2+\gamma_d^2}-\nonumber \\
&&- \frac{(\alpha+\beta)\cos(\alpha+\beta)t+\gamma_d\sin(\alpha+\beta)t}{(\alpha+\beta)^2+\gamma_d^2} \Bigg)+\nonumber \\
&&+\frac{\gamma_e \rho_{89}e^{-\gamma_e t}}{2}\Bigg(\frac{(\alpha-\beta)\sin(\alpha-\beta)t-\gamma_e\cos(\alpha-\beta)t}{(\alpha-\beta)^2+\gamma_e^2}+\nonumber \\
&&+\frac{(\alpha+\beta)\sin(\alpha+\beta)t-\gamma_e\cos(\alpha+\beta)t}{(\alpha+\beta)^2+\gamma_e^2}\bigg)+\nonumber \\
&&+\frac{i \gamma_e\rho_{69}e^{-\gamma_e t}}{2}\Bigg(\frac{(\alpha-\beta)\cos(\alpha-\beta)t+\gamma_e\sin(\alpha-\beta)t}{(\alpha-\beta)^2+\gamma_e^2}-\nonumber \\
&&-
\frac{(\alpha+\beta)\cos(\alpha+\beta)t+\gamma_e\sin(\alpha+\beta)t}{(\alpha+\beta)^2+\gamma_e^2}
\Bigg);\nonumber
\end{eqnarray}
\begin{eqnarray}
&&\int e^{\frac{\gamma_d+\gamma_e}{2}t}F_{23}(t)\sin \alpha t dt=\nonumber \\
&&=-\frac{\gamma_d \rho_{56}e^{-\gamma_d t}}{2}\Bigg(\frac{(\alpha-\beta)\cos(\alpha-\beta)t+\gamma_d\sin(\alpha-\beta)t}{(\alpha-\beta)^2+\gamma_d^2}+\nonumber \\
&&+\frac{(\alpha+\beta)\cos(\alpha+\beta)t+\gamma_d\sin(\alpha+\beta)t}{(\alpha+\beta)^2+\gamma_d^2}\bigg)-\nonumber \\
&&-\frac{i \gamma_d\rho_{58}e^{-\gamma_dt}}{2}\Bigg(\frac{(\alpha-\beta)\sin(\alpha-\beta)t-\gamma_d\cos(\alpha-\beta)t}{(\alpha-\beta)^2+\gamma_d^2}-\nonumber \\
&&- \frac{(\alpha+\beta)\sin(\alpha+\beta)t+\gamma_d\cos(\alpha+\beta)t}{(\alpha+\beta)^2+\gamma_d^2} \Bigg)-\nonumber \\
&&\frac{\gamma_e \rho_{89}e^{-\gamma_e t}}{2}\Bigg(\frac{(\alpha-\beta)\cos(\alpha-\beta)t+\gamma_e\sin(\alpha-\beta)t}{(\alpha-\beta)^2+\gamma_e^2}+\nonumber \\
&&+\frac{(\alpha+\beta)\cos(\alpha+\beta)t+\gamma_e\sin(\alpha+\beta)t}{(\alpha+\beta)^2+\gamma_e^2}\bigg)+\nonumber \\
&&+\frac{i \gamma_e\rho_{69}e^{-\gamma_e t}}{2}\Bigg(\frac{(\alpha-\beta)\sin(\alpha-\beta)t-\gamma_e\cos(\alpha-\beta)t}{(\alpha-\beta)^2+\gamma_e^2}-\nonumber \\
&&-
\frac{(\alpha+\beta)\sin(\alpha+\beta)t+\gamma_e\cos(\alpha+\beta)t}{(\alpha+\beta)^2+\gamma_e^2}
\Bigg)\nonumber \\
&&\gamma_{43}=\rho_{43}-\frac{i\gamma_d \rho_{56}}{2}\bigg(\frac{\alpha-\beta}{(\alpha-\beta)^2+\gamma_d^2} +\frac{\alpha+\beta}{(\alpha+\beta)^2+\gamma_d^2}\bigg)-\nonumber \\
&&-\frac{\gamma_d^2\rho_{58}}{2}\bigg(\frac{1}{(\alpha-\beta)^2+\gamma_d^2}+\frac{1}{(\alpha+\beta)^2+\gamma_d^2}\bigg)-\nonumber \\
&&-\frac{i\gamma_e \rho_{89}}{2}\bigg(\frac{\alpha-\beta}{(\alpha-\beta)^2+\gamma_e^2} +\frac{\alpha+\beta}{(\alpha+\beta)^2+\gamma_e^2}\bigg)+\nonumber \\&&+\frac{\gamma_e^2\rho_{69}}{2}\bigg(\frac{1}{(\alpha-\beta)^2+\gamma_e^2}+\frac{1}{(\alpha+\beta)^2+\gamma_e^2}\bigg);\nonumber \\
&&\gamma_{23}=\rho_{23}+\frac{\gamma_d^2 \rho_{56}}{2}\bigg(\frac{1}{(\alpha-\beta)^2+\gamma_d^2} +\frac{1}{(\alpha+\beta)^2+\gamma_d^2}\bigg)+\nonumber \\
&&+\frac{i\gamma_d\rho_{58}}{2}\bigg(\frac{\alpha-\beta}{(\alpha-\beta)^2+\gamma_d^2}-\frac{\alpha+\beta}{(\alpha+\beta)^2+\gamma_d^2}\bigg)+\nonumber \\
&&+\frac{\gamma_e^2
\rho_{89}}{2}\bigg(\frac{1}{(\alpha-\beta)^2+\gamma_e^2}
+\frac{1}{(\alpha+\beta)^2+\gamma_e^2}\bigg)-\nonumber
\\&&-\frac{i\gamma_e\rho_{69}}{2}\bigg(\frac{\alpha-\beta}{(\alpha-\beta)^2+\gamma_e^2}+\frac{\alpha+\beta}{(\alpha+\beta)^2+\gamma_e^2}\bigg);\nonumber
\end{eqnarray}
\begin{eqnarray}
&&\rho_{42}(t)=\big(\rho_{24}+\rho_{42}\big)e^{-\gamma_d t}-\bigg\{e^{-\gamma_d t}\big(\gamma_y \cos \alpha t-i \gamma_{x} \sin \alpha t\big)+\nonumber \\&&+i\gamma_d e^{-\gamma_d t}\big(\cos 2\alpha t \int e^{\gamma_d t}(\rho_{88}(t)-\rho_{66}(t))\sin 2 \alpha t dt-\nonumber \\
&&-\sin 2\alpha t \int e^{\gamma_d t}(\rho_{88}(t)-\rho_{66}(t))\cos 2 \alpha t dt\big)\bigg\};\nonumber\\
&&\rho_{41}(t)=\gamma_{41}e^{-\frac{\gamma_d}{2}t}\cos \alpha t+i\gamma_{21}e^{-\frac{\gamma_d}{2}t}\sin \alpha t+\nonumber \\
&&+e^{-\frac{\gamma_d}{2}t}\bigg\{\cos \alpha t \int e^{\frac{\gamma_d}{2}t}\big(F_{41} \cos \alpha t-iF_{21}\sin \alpha t\big)dt+\nonumber \\
&&+i\sin \alpha t \int e^{\frac{\gamma_d}{2}t}\big(F_{21} \cos \alpha t-iF_{41}\sin \alpha t\big)dt\bigg\};\nonumber\\
&&F_{21}(t)=\gamma_d \rho_{54}(t)+\gamma_e \rho_{87}(t); \nonumber \\
&&F_{41}(t)=\gamma_d \rho_{52}(t)+\gamma_e \rho_{63}(t); \nonumber \\
&&\int e^{\frac{\gamma_d t}{2}}\big(F_{21}\cos \alpha t-i F_{41} \sin \alpha t \big)dt=\nonumber\\
&&=\gamma_d \Bigg(\rho_{54}\frac{2\alpha \sin 2\alpha t-\gamma_{d}\cos 2 \alpha t}{4\alpha^2+\gamma_{d}^2}+\nonumber \\
&&+i\rho_{52}\frac{2\alpha \cos 2\alpha t+\gamma_{d}\sin 2 \alpha t}{4\alpha^2+\gamma_{d}^2}\Bigg)e^{-\gamma_d t}+\nonumber \\
&&+\frac{\gamma_e}{2}\Bigg(\rho_{87}\bigg(\frac{(\alpha+\beta)\sin(\alpha+\beta)t-\gamma_e \cos(\alpha+\beta)t}{(\alpha+\beta)^2+\gamma_e^2}+\nonumber\\
&&+\frac{(\alpha-\beta)\sin(\alpha-\beta)t-\gamma_e \cos(\alpha-\beta)t}{(\alpha-\beta)^2+\gamma_e^2}\bigg)+\nonumber \\
&&+i\rho_{67}\bigg(\frac{(\alpha-\beta)\cos(\alpha-\beta)t+\gamma_e \sin(\alpha-\beta)t}{(\alpha-\beta)^2+\gamma_e^2}-\nonumber \\
&&-\frac{(\alpha+\beta)\cos(\alpha+\beta)t+\gamma_e \sin(\alpha+\beta)t}{(\alpha+\beta)^2+\gamma_e^2}\bigg)+\nonumber\\
&&+i\rho_{63}\bigg(\frac{(\alpha-\beta)\cos(\alpha-\beta)t+\gamma_e \sin(\alpha-\beta)t}{(\alpha-\beta)^2+\gamma_e^2}+\nonumber \\
&&+\frac{(\alpha+\beta)\cos(\alpha+\beta)t+\gamma_e \sin(\alpha+\beta)t}{(\alpha+\beta)^2+\gamma_e^2}\bigg)+\nonumber\\
&&+\rho_{83}\bigg(\frac{(\alpha-\beta)\sin(\alpha-\beta)t-\gamma_e \cos(\alpha-\beta)t}{(\alpha-\beta)^2+\gamma_e^2}-\nonumber\\
&&-\frac{(\alpha+\beta)\sin(\alpha+\beta)t-\gamma_e \cos(\alpha+\beta)t}{(\alpha+\beta)^2+\gamma_e^2}\bigg)\Bigg)e^{-\gamma_e t};\nonumber
\end{eqnarray}
\begin{eqnarray}
&&\int e^{\frac{\gamma_d t}{2}}\big(F_{41}\cos \alpha t-i F_{21} \sin \alpha t \big)dt=\nonumber\\
&&=\gamma_d \Bigg(\rho_{52}\frac{2\alpha \sin 2\alpha t-\gamma_{d}\cos 2 \alpha t}{4\alpha^2+\gamma_{d}^2}+\nonumber \\
&&+i\rho_{54}\frac{2\alpha \cos 2\alpha t+\gamma_{d}\sin 2 \alpha t}{4\alpha^2+\gamma_{d}^2}\Bigg)e^{-\gamma_d t}+\nonumber \\
&&+\frac{\gamma_e}{2}\Bigg(\rho_{63}\bigg(\frac{(\alpha+\beta)\sin(\alpha+\beta)t-\gamma_e \cos(\alpha+\beta)t}{(\alpha+\beta)^2+\gamma_e^2}+\nonumber\\
&&+\frac{(\alpha-\beta)\sin(\alpha-\beta)t-\gamma_e \cos(\alpha-\beta)t}{(\alpha-\beta)^2+\gamma_e^2}\bigg)+\nonumber \\
&&+i\rho_{83}\bigg(\frac{(\alpha-\beta)\cos(\alpha-\beta)t+\gamma_e \sin(\alpha-\beta)t}{(\alpha-\beta)^2+\gamma_e^2}-\nonumber \\
&&-\frac{(\alpha+\beta)\cos(\alpha+\beta)t+\gamma_e \sin(\alpha+\beta)t}{(\alpha+\beta)^2+\gamma_e^2}\bigg)+\nonumber\\
&&+i\rho_{87}\bigg(\frac{(\alpha-\beta)\cos(\alpha-\beta)t+\gamma_e \sin(\alpha-\beta)t}{(\alpha-\beta)^2+\gamma_e^2}+\nonumber \\
&&+\frac{(\alpha+\beta)\cos(\alpha+\beta)t+\gamma_e \sin(\alpha+\beta)t}{(\alpha+\beta)^2+\gamma_e^2}\bigg)+\nonumber\\
&&+\rho_{67}\bigg(\frac{(\alpha-\beta)\sin(\alpha-\beta)t-\gamma_e \cos(\alpha-\beta)t}{(\alpha-\beta)^2+\gamma_e^2}-\nonumber\\
&&-\frac{(\alpha+\beta)\sin(\alpha+\beta)t-\gamma_e \cos(\alpha+\beta)t}{(\alpha+\beta)^2+\gamma_e^2}\bigg)\Bigg)e^{-\gamma_e t};\nonumber
\end{eqnarray}
\newline
\begin{eqnarray}
&& \gamma_{41}=\rho_{41}+\gamma_{d}\big(\frac{\gamma_{d}\rho_{52}-2 \alpha i \rho_{54}}{4 \alpha^2+\gamma_d^2}\big)+\nonumber \\
&&+\frac{\gamma_e \rho_{63}}{2}\Bigg(\frac{\gamma_e}{(\alpha-\beta)^2+\gamma_e^2}+\frac{\gamma_e}{(\alpha+\beta)^2+\gamma_e^2}\bigg)-\nonumber \\
&&-\frac{i\gamma_e \rho_{83}}{2}\Bigg(\frac{(\alpha-\beta)}{(\alpha-\beta)^2+\gamma_e^2}-\frac{(\alpha+\beta)}{(\alpha+\beta)^2+\gamma_e^2}\bigg)-\nonumber \\
&&-\frac{i\gamma_e \rho_{87}}{2}\Bigg(\frac{(\alpha-\beta)}{(\alpha-\beta)^2+\gamma_e^2}+\frac{(\alpha+\beta)}{(\alpha+\beta)^2+\gamma_e^2}\bigg)+\nonumber \\
&&+\frac{\gamma_e \rho_{67}}{2}\Bigg(\frac{\gamma_e}{(\alpha-\beta)^2+\gamma_e^2}-\frac{\gamma_e}{(\alpha+\beta)^2+\gamma_e^2}\bigg);\nonumber \\
&& \gamma_{21}=\rho_{21}+\gamma_{d}\big(\frac{\gamma_{d}\rho_{54}-2 \alpha i \rho_{52}}{4 \alpha^2+\gamma_d^2}\big)+\nonumber \\
&&+\frac{\gamma_e \rho_{87}}{2}\Bigg(\frac{\gamma_e}{(\alpha-\beta)^2+\gamma_e^2}+\frac{\gamma_e}{(\alpha+\beta)^2+\gamma_e^2}\bigg)-\nonumber \\
&&-\frac{i\gamma_e \rho_{67}}{2}\Bigg(\frac{(\alpha-\beta)}{(\alpha-\beta)^2+\gamma_e^2}-\frac{(\alpha+\beta)}{(\alpha+\beta)^2+\gamma_e^2}\bigg)-\nonumber \\
&&-\frac{i\gamma_e \rho_{63}}{2}\Bigg(\frac{(\alpha-\beta)}{(\alpha-\beta)^2+\gamma_e^2}+\frac{(\alpha+\beta)}{(\alpha+\beta)^2+\gamma_e^2}\bigg)+\nonumber \\
&&+\frac{\gamma_e \rho_{83}}{2}\Bigg(\frac{\gamma_e}{(\alpha-\beta)^2+\gamma_e^2}-\frac{\gamma_e}{(\alpha+\beta)^2+\gamma_e^2}\bigg);\nonumber
\end{eqnarray}
\begin{eqnarray}
&&\rho_{39}(t)=\rho_{39}e^{-\frac{3}{2}\gamma_{e}t};\nonumber \\
&&\rho_{38}(t)=\big(\rho_{38}\cos \beta t-i \rho_{36}\sin \beta t\big)e^{-(\frac{\gamma_d}{2}+\gamma_e)t};\nonumber \\
&&\rho_{37}(t)=\rho_{37}e^{-\gamma_{e}t};\nonumber \\
&&\rho_{36}(t)=\big(\rho_{36}\cos \beta t-i \rho_{38}\sin \beta t\big)e^{-(\frac{\gamma_d}{2}+\gamma_e)t};\nonumber \\
&&\rho_{35}(t)=\rho_{35}e^{-(\gamma_{d}+\frac{\gamma_e}{2})t};\nonumber \\
&&\rho_{34}(t)=e^{-\frac{\gamma_d+\gamma_e}{2}t}\Bigg(\gamma_{34}\cos \alpha t-i\gamma_{32}\sin \alpha t+\nonumber \\
&&+i \cos \alpha t \int e^{\frac{\gamma_d+\gamma_e}{2}t} F_{32}(t)\sin \alpha t dt-\nonumber \\
&&-i\sin \alpha t\int e^{\frac{\gamma_d+\gamma_e}{2}t} F_{32}(t)\cos \alpha t dt \Bigg);\nonumber\\
&&\rho_{33}(t)=\frac{1}{2}\bigg(2\rho_{77}+\rho_{66}+\rho_{88}+2\rho_{99}+\nonumber\\
 &&+\frac{\gamma_{d}^{2}\big(\rho_{66}-\rho_{88}\big)}{4\beta^2+\gamma_d^2}+\frac{2i\beta\gamma_{d}\big(\rho_{86}-\rho_{68}\big)}{4\beta^2+\gamma_d^2}\bigg)
 e^{-\gamma_{e}t}+ \nonumber \\
 &&+\frac{\gamma_d}{2}\big(\rho_{66}-\rho_{88}\big)\frac{2\beta\sin 2\beta t-\gamma_{d}\cos2 \beta t}{4\beta^2+\gamma_d^2}e^{-(\gamma_d+\gamma_e)t}-\nonumber \\
 &&-\frac{i\gamma_d}{2}\big(\rho_{86}-\rho_{68}\big)\frac{2\beta\cos 2\beta t+\gamma_{d}\sin2 \beta t}{4\beta^2+\gamma_d^2}e^{-(\gamma_d+\gamma_e)t}-\nonumber\\
 &&-\frac{1}{2}\big(\rho_{66}+\rho_{88}\big)e^{-(\gamma_e+\gamma_d)t}-\rho_{99}e^{-2\gamma_et};\nonumber
 \\
&&\rho_{32}(t)=e^{-\frac{\gamma_d+\gamma_e}{2}t}\Bigg(\gamma_{32}\cos \alpha t-i\gamma_{34}\sin \alpha t\big)+\nonumber \\
&&+ \cos \alpha t \int e^{\frac{\gamma_d+\gamma_e}{2}t} F_{32}(t)\cos \alpha t dt+\nonumber \\
&&+\sin \alpha t\int e^{\frac{\gamma_d+\gamma_e}{2}t} F_{32}(t)\sin \alpha t dt\Bigg); \nonumber\\
&&F_{32}(t)=\gamma_{d}\rho_{65}(t)+\gamma_{e}\rho_{98}(t);\nonumber\\
&&\int e^{\frac{\gamma_d+\gamma_e}{2}t}F_{32}(t)\sin \alpha t dt=\nonumber \\
&&=-\frac{\gamma_d \rho_{65}e^{-\gamma_d t}}{2}\Bigg(\frac{(\alpha-\beta)\cos(\alpha-\beta)t+\gamma_d\sin(\alpha-\beta)t}{(\alpha-\beta)^2+\gamma_d^2}+\nonumber \\
&&+\frac{(\alpha+\beta)\cos(\alpha+\beta)t+\gamma_d\sin(\alpha+\beta)t}{(\alpha+\beta)^2+\gamma_d^2}\bigg)+\nonumber \\
&&+\frac{i \gamma_d\rho_{85}e^{-\gamma_dt}}{2}\Bigg(\frac{(\alpha-\beta)\sin(\alpha-\beta)t-\gamma_d\cos(\alpha-\beta)t}{(\alpha-\beta)^2+\gamma_d^2}-\nonumber \\
&&- \frac{(\alpha+\beta)\sin(\alpha+\beta)t+\gamma_d\cos(\alpha+\beta)t}{(\alpha+\beta)^2+\gamma_d^2} \Bigg)-\nonumber \\
&&-\frac{\gamma_e \rho_{98}e^{-\gamma_e t}}{2}\Bigg(\frac{(\alpha-\beta)\cos(\alpha-\beta)t+\gamma_e\sin(\alpha-\beta)t}{(\alpha-\beta)^2+\gamma_e^2}+\nonumber \\
&&+\frac{(\alpha+\beta)\cos(\alpha+\beta)t+\gamma_e\sin(\alpha+\beta)t}{(\alpha+\beta)^2+\gamma_e^2}\bigg)-\nonumber \\
&&-\frac{i \gamma_e\rho_{96}e^{-\gamma_e t}}{2}\Bigg(\frac{(\alpha-\beta)\sin(\alpha-\beta)t-\gamma_e\cos(\alpha-\beta)t}{(\alpha-\beta)^2+\gamma_e^2}-\nonumber \\
&&- \frac{(\alpha+\beta)\sin(\alpha+\beta)t+\gamma_e\cos(\alpha+\beta)t}{(\alpha+\beta)^2+\gamma_e^2} \Big);\nonumber
\end{eqnarray}
\newpage
\begin{eqnarray}
&&\int e^{\frac{\gamma_d+\gamma_e}{2}t}F_{32}(t)\cos \alpha t dt=\nonumber \\
&&=\frac{\gamma_d \rho_{65}e^{-\gamma_d t}}{2}\Bigg(\frac{(\alpha-\beta)\sin(\alpha-\beta)t-\gamma_d\cos(\alpha-\beta)t}{(\alpha-\beta)^2+\gamma_d^2}+\nonumber \\
&&+\frac{(\alpha+\beta)\sin(\alpha+\beta)t-\gamma_d\cos(\alpha+\beta)t}{(\alpha+\beta)^2+\gamma_d^2}\bigg)+\nonumber \\
&&+\frac{i \gamma_d\rho_{85}e^{-\gamma_dt}}{2}\Bigg(\frac{(\alpha-\beta)\cos(\alpha-\beta)t+\gamma_d\sin(\alpha-\beta)t}{(\alpha-\beta)^2+\gamma_d^2}-\nonumber \\
&&- \frac{(\alpha+\beta)\cos(\alpha+\beta)t+\gamma_d\sin(\alpha+\beta)t}{(\alpha+\beta)^2+\gamma_d^2} \Bigg)+\nonumber \\
&&+\frac{\gamma_e \rho_{98}e^{-\gamma_e t}}{2}\Bigg(\frac{(\alpha-\beta)\sin(\alpha-\beta)t-\gamma_e\cos(\alpha-\beta)t}{(\alpha-\beta)^2+\gamma_e^2}+\nonumber \\
&&+\frac{(\alpha+\beta)\sin(\alpha+\beta)t-\gamma_e\cos(\alpha+\beta)t}{(\alpha+\beta)^2+\gamma_e^2}\bigg)+\nonumber \\
&&+\frac{i \gamma_e\rho_{96}e^{-\gamma_e t}}{2}\Bigg(\frac{(\alpha-\beta)\cos(\alpha-\beta)t+\gamma_e\sin(\alpha-\beta)t}{(\alpha-\beta)^2+\gamma_e^2}-\nonumber \\
&&- \frac{(\alpha+\beta)\cos(\alpha+\beta)t+\gamma_e\sin(\alpha+\beta)t}{(\alpha+\beta)^2+\gamma_e^2} \Big)\nonumber \\
&& \gamma_{34}=\rho_{34}+\frac{i\gamma_d \rho_{65}}{2}\big(\frac{\alpha-\beta}{(\alpha-\beta)^2+\gamma_d^2}+\frac{\alpha+\beta}{(\alpha+\beta)^2+\gamma_d^2}\big)-\nonumber \\
&&-\frac{\gamma_d \rho_{85}}{2}\big(\frac{\gamma_d}{(\alpha-\beta)^2+\gamma_d^2}+\frac{\gamma_d}{(\alpha+\beta)^2+\gamma_d^2}\big)+\nonumber \\
&&+\frac{i\gamma_e \rho_{98}}{2}\big(\frac{\alpha-\beta}{(\alpha-\beta)^2+\gamma_e^2}+\frac{\alpha+\beta}{(\alpha+\beta)^2+\gamma_e^2}\big)+\nonumber \\
&&+\frac{\gamma_e \rho_{96}}{2}\big(\frac{\gamma_e}{(\alpha-\beta)^2+\gamma_e^2}+\frac{\gamma_e}{(\alpha+\beta)^2+\gamma_e^2}\big);\nonumber \\
&& \gamma_{32}=\rho_{32}+\frac{\gamma_d \rho_{65}}{2}\big(\frac{\gamma_d}{(\alpha-\beta)^2+\gamma_d^2}+\frac{\gamma_d}{(\alpha+\beta)^2+\gamma_d^2}\big)-\nonumber \\
&&-\frac{i\gamma_d \rho_{85}}{2}\big(\frac{(\alpha-\beta)}{(\alpha-\beta)^2+\gamma_d^2}-\frac{(\alpha+\beta)}{(\alpha+\beta)^2+\gamma_d^2}\big)+\nonumber \\
&&+\frac{\gamma_e \rho_{98}}{2}\big(\frac{\gamma_e}{(\alpha-\beta)^2+\gamma_e^2}+\frac{\gamma_e}{(\alpha+\beta)^2+\gamma_e^2}\big)-\nonumber \\
&&-\frac{i\gamma_e \rho_{96}}{2}\big(\frac{(\alpha-\beta)}{(\alpha-\beta)^2+\gamma_e^2}-\frac{\alpha+\beta}{(\alpha+\beta)^2+\gamma_e^2}\big);\nonumber \\
&&\rho_{31}(t)=\gamma_{31}e^{-\frac{\gamma_e}{2}t}-\rho_{97}e^{-\frac{3}{2}\gamma_e t}+\nonumber \\
&&+\frac{\gamma_d e^{-(\gamma_d+\frac{\gamma_e}{2})t}}{2}\Bigg\{\rho_{64}\bigg(\frac{(\alpha-\beta)\sin(\alpha-\beta)t-\gamma_d \cos \alpha t}{(\alpha-\beta)^2+\gamma_d^2}+\nonumber\\
&&+\frac{(\alpha+\beta)\sin(\alpha+\beta)t-\gamma_d \cos (\alpha+\beta) t}{(\alpha+\beta)^2+\gamma_d^2}\bigg)+\nonumber \\
&&+i\rho_{84}\bigg(\frac{(\alpha-\beta)\cos(\alpha-\beta)t+\gamma_d \sin (\alpha-\beta) t}{(\alpha-\beta)^2+\gamma_d^2}-\nonumber\\
&&-\frac{(\alpha+\beta)\cos(\alpha+\beta)t+\gamma_d \sin (\alpha+\beta) t}{(\alpha+\beta)^2+\gamma_d^2}\bigg)+\nonumber \\
&&+i\rho_{62}\bigg(\frac{(\alpha-\beta)\cos(\alpha-\beta)t+\gamma_d \sin (\alpha-\beta) t}{(\alpha-\beta)^2+\gamma_d^2}+\nonumber\\
&&+\frac{(\alpha+\beta)\cos(\alpha+\beta)t+\gamma_d \sin
(\alpha+\beta) t}{(\alpha+\beta)^2+\gamma_d^2}\bigg)+\nonumber
\end{eqnarray}
\begin{eqnarray}
&&+\rho_{82}\bigg(\frac{(\alpha-\beta)\sin(\alpha-\beta)t-\gamma_d \cos \alpha t}{(\alpha-\beta)^2+\gamma_d^2}-\nonumber\\
&&-\frac{(\alpha+\beta)\sin(\alpha+\beta)t-\gamma_d \cos
(\alpha+\beta)t}{(\alpha+\beta)^2+\gamma_d^2}\bigg)\Bigg\};\nonumber \\
&&\gamma_{31}=\rho_{31}+\frac{\gamma_d^2}{2}(\rho_{64}+\rho_{82})\nonumber\\
&&\bigg(\frac{1}{(\alpha-\beta)^2+\gamma_d^2}+\frac{1}{(\alpha+\beta)^2+\gamma_d^2}\bigg)-\nonumber\\
&&-\frac{i\gamma_d(\rho_{84}+\rho_{62})}{2}\frac{(\alpha-\beta)}{(\alpha-\beta)^2+\gamma_d^2}+\nonumber\\
&&+\frac{i\gamma_d(\rho_{84}-\rho_{62})}{2}\frac{(\alpha+\beta)}{(\alpha+\beta)^2+\gamma_d^2};\nonumber \\
&&\rho_{29}(t)=e^{-(\frac{\gamma_d}{2}+\gamma_e)t}\big(\rho_{29}\cos \alpha t-i\rho_{49}\sin \alpha t\big);\nonumber\\
&&\rho_{28}(t)=e^{-(\frac{\gamma_e}{2}+\gamma_d)t}\big(\cos \alpha t(\rho_{28}\cos \beta t-i \rho_{26} \sin \beta t)+ \nonumber \\
&&+\sin \alpha t(i \rho_{48}\cos \beta t+ \rho_{46} \sin \beta t\big);\nonumber \\
&&\rho_{27}(t)=e^{-\frac{\gamma_d+\gamma_e}{2}t}\big(\gamma_{27}\cos \alpha t+i \gamma_{47}\sin \alpha t\big)-\nonumber\\
    &&-ie^{-\frac{\gamma_d+\gamma_e}{2}t}\big(\cos \alpha t\int e^{\frac{\gamma_d+\gamma_e}{2}t}F_{47}(t)\sin \alpha t-\nonumber \\
    &&-\sin \alpha t\int e^{\frac{\gamma_d+\gamma_e}{2}t}F_{47}(t)\cos \alpha t\big);\nonumber \\
&&\rho_{26}(t)=e^{-(\frac{\gamma_e}{2}+\gamma_d)t}\big(\sin \alpha t(i \rho_{46}\cos  \beta t+ \rho_{48} \sin \beta t)+ \nonumber \\
&&+ \cos \beta t(\rho_{26} \cos \beta t-i \rho_{28} \sin \beta t)\big);\nonumber \\
 &&\rho_{25}(t)=e^{-\frac{3}{2}\gamma_d t}(\rho_{25}\cos \alpha t+i\rho_{45}\sin \alpha t);\nonumber\\
 && \rho_{24}(t)=\big(\rho_{24}+\rho_{42}\big)e^{-\gamma_d t}+\bigg\{e^{-\gamma_d t}\big(\gamma_y \cos \alpha t-i \gamma_{x} \sin \alpha t\big)+\nonumber \\&&+i\gamma_d e^{-\gamma_d t}\big(\cos 2\alpha t \int e^{\gamma_d t}(\rho_{88}(t)-\rho_{66}(t))\sin 2 \alpha t dt-\nonumber \\
&&-\sin 2\alpha t \int e^{\gamma_d t}(\rho_{88}(t)-\rho_{66}(t))\cos 2 \alpha t dt\big)\bigg\};\nonumber\\
&&\rho_{23}(t)=e^{-\frac{\gamma_d+\gamma_e}{2}t}\Bigg\{\gamma_{23}\cos \alpha t+i\gamma_{43}\sin \alpha t+\nonumber \\
&&+\cos \alpha t \int e^{\frac{\gamma_d+\gamma_e}{2}t}F_{23}(t)\cos \alpha t dt+\nonumber\\
&&+\sin \alpha t \int e^{\frac{\gamma_d+\gamma_e}{2}t}F_{23}(t)\sin \alpha t dt\Bigg\};\nonumber \\
&&\rho_{22}(t)=\frac{1}{2}\big(\rho_{22}+\rho_{44}+2\rho_{55}+\rho_{88}+\rho_{66}\big)e^{-\gamma_d t}-\nonumber \\
 &&-\rho_{55}e^{-2\gamma_d t}-\frac{1}{2}(\rho_{88}+\rho_{66})e^{-(\gamma_d+\gamma_e) t}+\nonumber \\
 &&+\frac{e^{-\gamma_d t}}{2}\big(\gamma_x \cos 2\alpha t-i\gamma_y \sin 2\alpha t\big)+\nonumber \\
 &&+\frac{\gamma_d e^{-\gamma_d t}}{2}\big(\cos 2\alpha t \int e^{\gamma_d t}(\rho_{88}(t)-\rho_{66}(t))\cos 2\alpha t dt+\nonumber \\
 &&+\sin 2\alpha t \int e^{\gamma_d t}(\rho_{88}(t)-\rho_{66}(t))\sin 2\alpha t dt\big);\nonumber \\
 &&\rho_{21}(t)=\gamma_{21}e^{-\frac{\gamma_d}{2}t}\cos \alpha t+i\gamma_{41}e^{-\frac{\gamma_d}{2}t}\sin \alpha t+\nonumber \\
&&+e^{-\frac{\gamma_d}{2}t}\bigg\{\cos \alpha t \int e^{\frac{\gamma_d}{2}t}\big(F_{21} \cos \alpha t-iF_{41}\sin \alpha t\big)dt+\nonumber \\
&&+i\sin \alpha t \int e^{\frac{\gamma_d}{2}t}\big(F_{41} \cos \alpha t-iF_{21}\sin \alpha t\big)dt\bigg\};\nonumber
\end{eqnarray}
\begin{eqnarray}
&&\rho_{19}(t)=\rho_{19}e^{-\gamma_e t};\nonumber \\
&&\rho_{18}(t)=e^{-\frac{\gamma_d+\gamma_e}{2}t}\big(\rho_{18} \cos \beta t-i\rho_{16} \sin \beta t\big);\nonumber \\
&&\rho_{17}(t)=\gamma_{17}e^{-\frac{\gamma_e}{2}t}-\rho_{39}e^{-\frac{3}{2}\gamma_e t}+\nonumber \\
&&+\frac{\gamma_d e^{-(\gamma_d+\frac{\gamma_e}{2})t}}{2}\Bigg\{\rho_{28}\bigg(\frac{(\alpha-\beta)\sin(\alpha-\beta)t-\gamma_d \cos \alpha t}{(\alpha-\beta)^2+\gamma_d^2}+\nonumber\\
&&+\frac{(\alpha+\beta)\sin(\alpha+\beta)t-\gamma_d \cos (\alpha+\beta) t}{(\alpha+\beta)^2+\gamma_d^2}\bigg)-\nonumber \\
&&-i\rho_{26}\bigg(\frac{(\alpha-\beta)\cos(\alpha-\beta)t+\gamma_d \sin (\alpha-\beta) t}{(\alpha-\beta)^2+\gamma_d^2}-\nonumber\\
&&-\frac{(\alpha+\beta)\cos(\alpha+\beta)t+\gamma_d \sin (\alpha+\beta) t}{(\alpha+\beta)^2+\gamma_d^2}\bigg)-\nonumber \\
&&-i\rho_{48}\bigg(\frac{(\alpha-\beta)\cos(\alpha-\beta)t+\gamma_d \sin (\alpha-\beta) t}{(\alpha-\beta)^2+\gamma_d^2}+\nonumber\\
&&+\frac{(\alpha+\beta)\cos(\alpha+\beta)t+\gamma_d \sin (\alpha+\beta) t}{(\alpha+\beta)^2+\gamma_d^2}\bigg)+\nonumber \\
&&+\rho_{46}\bigg(\frac{(\alpha-\beta)\sin(\alpha-\beta)t-\gamma_d \cos \alpha t}{(\alpha-\beta)^2+\gamma_d^2}-\nonumber\\
&&-\frac{(\alpha+\beta)\sin(\alpha+\beta)t-\gamma_d \cos (\alpha+\beta) t}{(\alpha+\beta)^2+\gamma_d^2}\bigg)\Bigg\};\nonumber \\
&&\gamma_{17}=\rho_{17}+\frac{\gamma_d^2}{2}(\rho_{46}+\rho_{28})\nonumber\\
&&\bigg(\frac{1}{(\alpha-\beta)^2+\gamma_d^2}+\frac{1}{(\alpha+\beta)^2+\gamma_d^2}\bigg)+\nonumber\\
&&+\frac{i\gamma_d(\rho_{48}+\rho_{26})}{2}\frac{(\alpha-\beta)}{(\alpha-\beta)^2+\gamma_d^2}+ \nonumber\\
&&+\frac{i\gamma_d(\rho_{48}-\rho_{26})}{2}\frac{(\alpha+\beta)}{(\alpha+\beta)^2+\gamma_d^2};\nonumber \\
&&\rho_{16}(t)=e^{-\frac{\gamma_d+\gamma_e}{2}t}\big(\rho_{16} \cos \beta t-i\rho_{18} \sin \beta t\big);\nonumber \\
&&\rho_{15}(t)=\rho_{15}e^{-\gamma_d t}; \nonumber \\
&& \rho_{14}(t)=e^{-\frac{\gamma_d}{2}t}\Bigg\{\gamma_{14}\cos \alpha t-i \gamma_{12}\sin \alpha t+ \nonumber \\
&&+\cos \alpha t\int e^{\frac{\gamma_d}{2}t}(F_{14}(t)\cos \alpha t+i F_{12}(t)\sin \alpha t)dt- \nonumber \\
&&-i \sin \alpha t \int e^{\frac{\gamma_d}{2}t}(F_{12}(t)\cos \alpha t+i F_{14}(t)\sin \alpha t)dt\Bigg\};\nonumber \\
&&\rho_{12}(t)=e^{-\frac{\gamma_d}{2}t}\Bigg\{\gamma_{12}\cos \alpha t-i \gamma_{14}\sin \alpha t+ \nonumber \\
&&+\cos \alpha t\int e^{\frac{\gamma_d}{2}t}(F_{12}(t)\cos \alpha t+i F_{14}(t)\sin \alpha t)dt- \nonumber \\
&&-i \sin \alpha t \int e^{\frac{\gamma_d}{2}t}(F_{14}(t)\cos \alpha t+i F_{12}(t)\sin \alpha t)dt\Bigg\};\nonumber \\
&&F_{12}(t)=\gamma_d\rho_{45}(t)+\gamma_e \rho_{78}(t); \nonumber \\
&&F_{14}(t)=\gamma_d\rho_{25}(t)+\gamma_e \rho_{36}(t); \nonumber
\end{eqnarray}
\begin{eqnarray}
&&\int e^{\frac{\gamma_d t}{2}}\big(F_{12}\cos \alpha t+i F_{14} \sin \alpha t \big)dt=\nonumber\\
&&=\gamma_d \Bigg(\rho_{45}\frac{2\alpha \sin 2\alpha t-\gamma_{d}\cos 2 \alpha t}{4\alpha^2+\gamma_{d}^2}-\nonumber \\
&&-i\rho_{25}\frac{2\alpha \cos 2\alpha t+\gamma_{d}\sin 2 \alpha t}{4\alpha^2+\gamma_{d}^2}\Bigg)e^{-\gamma_d t}+\nonumber \\
&&+\frac{\gamma_e}{2}\Bigg(\rho_{78}\bigg(\frac{(\alpha+\beta)\sin(\alpha+\beta)t-\gamma_e \cos(\alpha+\beta)t}{(\alpha+\beta)^2+\gamma_e^2}+\nonumber\\
&&+\frac{(\alpha-\beta)\sin(\alpha-\beta)t-\gamma_e \cos(\alpha-\beta)t}{(\alpha-\beta)^2+\gamma_e^2}\bigg)+\nonumber \\
&&-i\rho_{76}\bigg(\frac{(\alpha-\beta)\cos(\alpha-\beta)t+\gamma_e \sin(\alpha-\beta)t}{(\alpha-\beta)^2+\gamma_e^2}-\nonumber \\
&&-\frac{(\alpha+\beta)\cos(\alpha+\beta)t+\gamma_e \sin(\alpha+\beta)t}{(\alpha+\beta)^2+\gamma_e^2}\bigg)+\nonumber\\
&&-i\rho_{36}\bigg(\frac{(\alpha-\beta)\cos(\alpha-\beta)t+\gamma_e \sin(\alpha-\beta)t}{(\alpha-\beta)^2+\gamma_e^2}+\nonumber \\
&&+\frac{(\alpha+\beta)\cos(\alpha+\beta)t+\gamma_e \sin(\alpha+\beta)t}{(\alpha+\beta)^2+\gamma_e^2}\bigg)+\nonumber\\
&&+\rho_{38}\bigg(\frac{(\alpha-\beta)\sin(\alpha-\beta)t-\gamma_e \cos(\alpha-\beta)t}{(\alpha-\beta)^2+\gamma_e^2}-\nonumber\\
&&-\frac{(\alpha+\beta)\sin(\alpha+\beta)t-\gamma_e \cos(\alpha+\beta)t}{(\alpha+\beta)^2+\gamma_e^2}\bigg)\Bigg)e^{-\gamma_e t};\nonumber \\
&&\int e^{\frac{\gamma_d t}{2}}\big(F_{14}\cos \alpha t+i F_{12} \sin \alpha t \big)dt=\nonumber\\
&&=\gamma_d \Bigg(\rho_{25}\frac{2\alpha \sin 2\alpha t-\gamma_{d}\cos 2 \alpha t}{4\alpha^2+\gamma_{d}^2}-\nonumber \\
&&-i\rho_{45}\frac{2\alpha \cos 2\alpha t+\gamma_{d}\sin 2 \alpha t}{4\alpha^2+\gamma_{d}^2}\Bigg)e^{-\gamma_d t}+\nonumber \\
&&+\frac{\gamma_e}{2}\Bigg(\rho_{36}\bigg(\frac{(\alpha+\beta)\sin(\alpha+\beta)t-\gamma_e \cos(\alpha+\beta)t}{(\alpha+\beta)^2+\gamma_e^2}+\nonumber\\
&&+\frac{(\alpha-\beta)\sin(\alpha-\beta)t-\gamma_e \cos(\alpha-\beta)t}{(\alpha-\beta)^2+\gamma_e^2}\bigg)+\nonumber \\
&&-i\rho_{38}\bigg(\frac{(\alpha-\beta)\cos(\alpha-\beta)t+\gamma_e \sin(\alpha-\beta)t}{(\alpha-\beta)^2+\gamma_e^2}-\nonumber \\
&&-\frac{(\alpha+\beta)\cos(\alpha+\beta)t+\gamma_e \sin(\alpha+\beta)t}{(\alpha+\beta)^2+\gamma_e^2}\bigg)+\nonumber\\
&&-i\rho_{78}\bigg(\frac{(\alpha-\beta)\cos(\alpha-\beta)t+\gamma_e \sin(\alpha-\beta)t}{(\alpha-\beta)^2+\gamma_e^2}+\nonumber \\
&&+\frac{(\alpha+\beta)\cos(\alpha+\beta)t+\gamma_e \sin(\alpha+\beta)t}{(\alpha+\beta)^2+\gamma_e^2}\bigg)+\nonumber\\
&&+\rho_{76}\bigg(\frac{(\alpha-\beta)\sin(\alpha-\beta)t-\gamma_e \cos(\alpha-\beta)t}{(\alpha-\beta)^2+\gamma_e^2}-\nonumber\\
&&-\frac{(\alpha+\beta)\sin(\alpha+\beta)t-\gamma_e \cos(\alpha+\beta)t}{(\alpha+\beta)^2+\gamma_e^2}\bigg)\Bigg)e^{-\gamma_e t};\nonumber
\end{eqnarray}
\begin{eqnarray}
&&\gamma_{12}=\rho_{12}+\gamma_{d}\big(\frac{\gamma_{d}\rho_{45}+2 \alpha i \rho_{25}}{4 \alpha^2+\gamma_d^2}\big)+\nonumber \\
&&+\frac{\gamma_e \rho_{78}}{2}\Bigg(\frac{\gamma_e}{(\alpha-\beta)^2+\gamma_e^2}+\frac{\gamma_e}{(\alpha+\beta)^2+\gamma_e^2}\bigg)+\nonumber \\
&&+\frac{i\gamma_e \rho_{76}}{2}\Bigg(\frac{(\alpha-\beta)}{(\alpha-\beta)^2+\gamma_e^2}-\frac{(\alpha+\beta)}{(\alpha+\beta)^2+\gamma_e^2}\bigg)-\nonumber \\
&&+\frac{i\gamma_e \rho_{36}}{2}\Bigg(\frac{(\alpha-\beta)}{(\alpha-\beta)^2+\gamma_e^2}+\frac{(\alpha+\beta)}{(\alpha+\beta)^2+\gamma_e^2}\bigg)+\nonumber \\
&&+\frac{\gamma_e \rho_{38}}{2}\Bigg(\frac{\gamma_e}{(\alpha-\beta)^2+\gamma_e^2}-\frac{\gamma_e}{(\alpha+\beta)^2+\gamma_e^2}\bigg);\nonumber \\
&& \gamma_{14}=\rho_{14}+\gamma_{d}\big(\frac{\gamma_{d}\rho_{25}+2 \alpha i \rho_{45}}{4 \alpha^2+\gamma_d^2}\big)+\nonumber \\
&&+\frac{\gamma_e \rho_{36}}{2}\Bigg(\frac{\gamma_e}{(\alpha-\beta)^2+\gamma_e^2}+\frac{\gamma_e}{(\alpha+\beta)^2+\gamma_e^2}\bigg)+\nonumber \\
&&+\frac{i\gamma_e \rho_{38}}{2}\Bigg(\frac{(\alpha-\beta)}{(\alpha-\beta)^2+\gamma_e^2}-\frac{(\alpha+\beta)}{(\alpha+\beta)^2+\gamma_e^2}\bigg)+\nonumber \\
&&+\frac{i\gamma_e \rho_{78}}{2}\Bigg(\frac{(\alpha-\beta)}{(\alpha-\beta)^2+\gamma_e^2}+\frac{(\alpha+\beta)}{(\alpha+\beta)^2+\gamma_e^2}\bigg)+\nonumber \\
&&+\frac{\gamma_e \rho_{76}}{2}\Bigg(\frac{\gamma_e}{(\alpha-\beta)^2+\gamma_e^2}-\frac{\gamma_e}{(\alpha+\beta)^2+\gamma_e^2}\bigg);\nonumber \\
&&\rho_{13}(t)=\gamma_{13}e^{-\frac{\gamma_e}{2}t}-\rho_{79}e^{-\frac{3\gamma_{e}}{2}t}+\nonumber \\
&&+\frac{\gamma_d }{2}e^{-\frac{(2\gamma_d+\gamma_e)}{2}t}\Bigg\{\rho_{46}\bigg(\frac{(\alpha-\beta)\sin(\alpha-\beta)t-\gamma_d\cos(\alpha-\beta)t}{(\alpha-\beta)^2+\gamma_d^2}+\nonumber\\
&&+\frac{(\alpha+\beta)\sin(\alpha+\beta)t-\gamma_d\cos(\alpha+\beta)t}{(\alpha+\beta)^2+\gamma_d^2}\bigg)-\nonumber\\
&&-i\rho_{48}\bigg(\frac{(\alpha-\beta)\cos(\alpha-\beta)t+\gamma_d\sin(\alpha-\beta)t}{(\alpha-\beta)^2+\gamma_d^2}-\nonumber\\
&&-\frac{(\alpha+\beta)\cos(\alpha+\beta)t+\gamma_d\sin(\alpha+\beta)t}{(\alpha+\beta)^2+\gamma_d^2}\bigg)-\nonumber\\
&&-i\rho_{26}\bigg(\frac{(\alpha-\beta)\cos(\alpha-\beta)t+\gamma_d\sin(\alpha-\beta)t}{(\alpha-\beta)^2+\gamma_d^2}+\nonumber\\
&&-\frac{(\alpha+\beta)\cos(\alpha+\beta)t+\gamma_d\sin(\alpha+\beta)t}{(\alpha+\beta)^2+\gamma_d^2}\bigg)+\nonumber\\
&&+\rho_{28}\bigg(\frac{(\alpha-\beta)\sin(\alpha-\beta)t-\gamma_d\cos(\alpha-\beta)t}{(\alpha-\beta)^2-\gamma_d^2}+\nonumber\\
&&-\frac{(\alpha+\beta)\sin(\alpha+\beta)t-\gamma_d\cos(\alpha+\beta)t}{(\alpha+\beta)^2+\gamma_d^2}\bigg)\Bigg\};\nonumber \\
&& \gamma_{13}=\rho_{13}+\rho_{79}+\nonumber \\
&&+\frac{\gamma_d \rho_{46}}{2}\Bigg(\frac{\gamma_d}{(\alpha-\beta)^2+\gamma_d^2}+\frac{\gamma_d}{(\alpha+\beta)^2+\gamma_d^2}\bigg)+\nonumber \\
&&+\frac{i\gamma_d \rho_{48}}{2}\Bigg(\frac{(\alpha-\beta)}{(\alpha-\beta)^2+\gamma_d^2}-\frac{(\alpha+\beta)}{(\alpha+\beta)^2+\gamma_d^2}\bigg)+\nonumber \\
&&+\frac{i\gamma_d \rho_{26}}{2}\Bigg(\frac{(\alpha-\beta)}{(\alpha-\beta)^2+\gamma_d^2}+\frac{(\alpha+\beta)}{(\alpha+\beta)^2+\gamma_d^2}\bigg)+\nonumber \\
&&+\frac{\gamma_d
\rho_{28}}{2}\Bigg(\frac{\gamma_d}{(\alpha-\beta)^2+\gamma_d^2}-\frac{\gamma_d}{(\alpha+\beta)^2+\gamma_d^2}\bigg);\nonumber
\end{eqnarray}
\begin{eqnarray}
&&\rho_{11}(t)=\rho_{11}+\rho_{22}+\rho_{33}+\rho_{44}+\rho_{55}+\rho_{66}+\rho_{77}+\rho_{88}+\rho_{99}-\nonumber \\
&&-(\rho_{22}+\rho_{44}+2\rho_{55}+\rho_{88}+\rho_{66})e^{-\gamma_d t}+\nonumber \\
&&+\rho_{55}e^{-2\gamma_d t}+(\rho_{66}+\rho_{88})e^{-(\gamma_d+\gamma_e)t}-\nonumber \\
&&-((\rho_{33}+\rho_{77}+2\rho_{99}+\rho_{88}+\rho_{66})e^{-\gamma_e t})+\nonumber \\
&&+\rho_{99}e^{-2\gamma_e t};\nonumber
\end{eqnarray}


\begin{thebibliography}{38}
\bibitem{r1} J.~Eisert, M.~B.~Plenio, S.~Bose, and J.~Hartley, Phys. Rev. Lett. {\bf 93}, 190402 (2004).
\bibitem{r2} M.~B.~Plenio, J.~Hartley and J.~Eisert, New J. Phys. {\bf 6}, 36 (2004).
\bibitem{r3} L.~Novotny and B.~Hecht, {\it Principles of Nano-optics}, Cambridge University Press, 2008, 939 p.
\bibitem{r4} P.~Schleich, {\it Quantum Optics in Phase Space}, Wiley-VCH, Berlin, 2001, 760 p.
\bibitem{r5} T.~Aoki, B.~Dayan, E.~D.~Wilcut, W.~P.~Bowen, A.~S.~Parkins, T.~J.~Kippenberg, K.~J.~Vahala, and H.~J.~Kimble, Nature {\bf 443},  671 (2006).
\bibitem{r6} L.~Allen, and J.~H.~Eberly, {\it Optical Resonance and two-Level Atoms} John Wiley and Sons, 1975.
\bibitem{r7} H.~I.~Yoo and J.~H.~Eberly, Phys. Rep. {\bf118}, 5, 240 (1985).
\bibitem{r8} M.~A.~Nielsen and I.~L.~Chuang, {\it Quantum Computation and Quantum Information} (Cambridge University Press, Cambridge, 2000).
\bibitem{r9} J.~J.~L.~Morton {\it et. al.}, Nature {\bf 455}, 1085 (2008).
\bibitem{r10} K.~Saeedi {\it et. al.} Science {\bf 342}, 830 (2013).
\bibitem{r11} J.~J.~Cirac and P.~Zoller, Phys. Rev. Lett {\bf 74}, 4094 (1993).
\bibitem{r12} D.~F.~V.~James, Appl. Phys B, {\bf 66,} 181 1998.
\bibitem{r13} D.~Leibfried, R.~Blatt, C.~Monroe, and D.~Wineland, Rev. Mod. Phys, {\bf75}, 281 (2003).
\bibitem{r14} H.~H\"{a}ffner , C.~F.~Roos, and R.~Blatt, Phys. Rep., {\bf469}, 4 155 (2008).
\bibitem{r15} P.~Treutlein, D.~Hunger, S.~Camerer, T.~W.~H\"{a}nsch, and J.~Reichel, Phys. Rev. Lett. {\bf99}, 140403 (2007).
\bibitem{r16} P.~C.~Maurer, {\it et. al.}, Nat. Phys. {\bf6}, 912 (2010).
\bibitem{r17} M.~V.~Gurudev Dutt, {\it et. al.}, Science {\bf316}, 1312 (2007).
\bibitem{r18} J.~Wrachtrup and F.~Jelezko, J.~Phys. J. Phys. Condens. Matter {\bf18}, S807 (2006).
\bibitem{r19} P.~Rabl, P.~Cappellaro, M.~V.~Gurudev Dutt, L.~Jiang, J.~R.~Maze, and M.~D.~Lukin, Phys. Rev. B {\bf79}, 041302 (R) (2009).
\bibitem{r20} Li-gong Zhou, L.~F.~Wei, M.~Gao, and X.~Wang, Phys. Rev. A {\bf81}, 042323 (2010).
\bibitem{r21} O.~Arcizet, V.~Jacques, A.~Siria, P.~Poncharal, P.~Vincent, and S.~Seidelin, Nat. Phys. {\bf7,} 879 (2011).
\bibitem{r22} D.~Rugar, R.~Budakian, H.~J.~Mamin, and B.~W.~Chui, Nature (London) {\bf430}, 15 (2004).
\bibitem{r23} P.~Treutlein, Science {\bf335}, 1584 (2012).
\bibitem{r24} S.~Kolkowitz, A.~C.~Bleszynski-Jayich, Q.~P.~Unterreithmeier, S.~D.~Bennett, P.~Rabl, J.~G.~E.~Harris, and M.~D.~Lukin, Science {\bf335}, 1603 (2012).
\bibitem{r25} Z.~-H.~Wang and V.~V.~Dobrovitski, Phys. Rev. B {\bf84}, 045303 (2011).
\bibitem{r26} L.~Chotorlishvili, D.~Sander, A.~Sukhov, V.~Dugaev, V.~R.~Vieira, A.~Komnik, and J.~Berakdar, Phys. Rev. B {\bf88}, 085201 (2013).
\bibitem{r26-1}S.~K.~Mishra, L.~Chotorlishvili, A.~R.~P.~Rau, and J.~Berakdar Phys. Rev. A {\bf90} 033817 (2014).
\bibitem{r27} G.~Chawla and S.~D.~Solares, Meas. Sci. Technol. {\bf20}, 015501 (2009).
\bibitem{r28} S.~D.~Solares and G.~Chawla, Meas. Sci. Technol. {\bf19}, 055502 (2008).
\bibitem{r30} R.~P.~Feynman {\it Statistical Mechanics}, Massachusets 1972.
\bibitem{r31} G.~Vidal, R.~F.~Werner, Phys. Rev.A {\bf65}, 032314 (2002).
\bibitem{r32} R.~Garcia and E.~T.~Herruzo, Nature Nanotechnology {\bf7}, 217 (2012).
\bibitem{r33} J.~R.~Lozano and R.~Garcia, Phys. Rev. Lett. {\bf100}, 076102 (2008).
\bibitem{r34} N.~N.~Bogolubov Jr, F.~L.~Kien, and A.~S.~Shumovski, Phys. Lett. {\bf101A}, 4, 201, (1984).
\bibitem{r35} H. Fr\"{o}hlich, Proc. R. Soc. London, Ser. A {\bf215}, 291 (1952); P.~Tarasewicz and D.~Baran, Phys. Rev. B {\bf73}, 094524 (2006); C. Kittel,
{\it Quantum Theory of Solids} (Wiley, New York,1987).
\bibitem{r36} D.~Saya, K.~Fukushima, H.~Toshiyoshi, G.~Hashiguchi, H.~Fujita, and H.~Kawakatsu, Sensors and
Actuators A {\bf95},  281 (2002).
\bibitem{r37} D.~Sarid, {\it Exploring Scanning Probe Microscopy with Mathematica} (Wiley, New York, 1997).
\bibitem{r38} Mazhar Ali, J. Phys. B: At. Mol. Opt. Phys. {\bf43}, 045504 (2010).
\bibitem{r39}  L. Derkacz  and L. Jak\'{o}bczyk,  Phys. Rev. A
{\bf74}, 032313 (2006).
\end{thebibliography}
\end{document}